%% file: main.tex
\documentclass[sigconf, nonacm]{acmart}

\usepackage[ruled,linesnumbered,vlined]{algorithm2e}

\newcommand\vldbdoi{XX.XX/XXX.XX}

\newcommand\vldbavailabilityurl{URL_TO_YOUR_ARTIFACTS}

\usepackage{enumitem}
\usepackage{threeparttable}
\usepackage{multirow}
\usepackage{tabularx}
\usepackage[linesnumbered,ruled,vlined]{algorithm2e}
\usepackage{graphicx}
\usepackage{subcaption}

\newcommand{\algcommentline}[1]{\noindent\textcolor{blue}{$\triangleright$~#1}\par}
\newcommand{\algcommentlineComment}[1]{\noindent\textcolor{gray}{$\triangleright$~#1}\par}

\begin{document}
\title{Leveraging the Spatial Hierarchy: Coarse-to-fine Trajectory Generation via Cascaded Hybrid Diffusion}

\author{Baoshen Guo}
\affiliation{%
  \institution{Singapore-MIT Alliance for Research and Technology}
  \city{Singapore}
  \state{Singapore}
}
\email{baoshen.guo@smart.mit.edu}

\author{Zhiqing Hong}
\affiliation{%
  \institution{Rutgers University}
  \city{Piscataway}
  \state{USA}
}
\email{zhiqing.hong@rutgers.edu}

\author{Junyi Li}
\affiliation{%
  \institution{Singapore-MIT Alliance for Research and Technology}
  \city{Singapore}
  \state{Singapore}
}
\email{junyi.li@smart.mit.edu}

\author{Shenhao Wang}
\affiliation{%
  \institution{University of Florida}
  \city{Gainesville}
  \state{USA}
}
\email{shenhaowang@ufl.edu}

\author{Jinhua Zhao}
\affiliation{%
  \institution{Massachusetts Institute of Technology}
  \city{Cambridge}
  \state{USA}
}
\email{jinhua@mit.edu}

\begin{abstract}
Urban mobility data has significant connections with economic growth and plays an essential role in various smart-city applications.
However, due to privacy concerns and substantial data collection costs, fine-grained human mobility trajectories are challenging to become publicly available on a large scale. 
A promising solution to address this issue is trajectory synthesizing, which generates synthetic trajectories that preserve aggregate spatiotemporal distributions while mitigating privacy risks. 
However, existing works often ignore the inherent structural complexity of trajectories, thus unable to handle complicated high-dimensional distributions and generate realistic fine-grained trajectories. 
In this paper, 
we propose \textit{Cardiff}, a coarse-to-fine \underline{\textbf{Ca}}scaded hyb\underline{\textbf{r}}id \underline{\textbf{diff}}usion-based trajectory synthesizing framework for fine-grained and privacy-preserving mobility generation.  
By leveraging the hierarchical nature of urban mobility, 
\textit{Cardiff} decomposes the generation process into two distinct levels, i.e., discrete road segment-level and continuous fine-grained GPS-level:
(i) At the segment level, 
to reduce computational costs and redundancy in raw trajectories, we first encode the discrete road segments into low-dimensional latent embeddings and design a diffusion transformer-based latent denoising network for segment-level trajectory synthesis. 
(ii) Taking the first stage of generation as conditions, we then design a fine-grained GPS-level conditional denoising network with a noise augmentation mechanism to achieve road-network-constrained and fine-grained generation. 
Additionally, the \textit{Cardiff} framework not only progressively generates high-fidelity trajectories through cascaded denoising but also flexibly enables a tunable balance between privacy preservation and utility.
Experimental results on three large real-world trajectory datasets demonstrate that our method outperforms state-of-the-art baselines in various metrics.

\end{abstract}

\maketitle


\ifdefempty{\vldbavailabilityurl}{}{
\vspace{.1cm}
\begingroup\small\noindent\raggedright\textbf{PVLDB Artifact Availability:}\\
The source code and other artifacts are available at \url{https://github.com/urban-mobility-generation/Cardiff}.
\endgroup
}

\input{Sec1-intro}
\input{Sec2-Problem}

\input{Sec3-method}

\input{Sec4-evaluation}

\input{Sec5-relatedworks}

\input{Sec6-discussion}

\input{Sec7-conclusion}




\bibliographystyle{ACM-Reference-Format}
\bibliography{sample}

\end{document}

%% file: Sec1-intro.tex
\section{Introduction}

Urban mobility trajectories play a crucial role in the development of smart cities by establishing dynamic connections between urban regions and infrastructures. These trajectories support a wide range of spatiotemporal applications, spanning from macro-level tasks such as traffic assignment~\cite{wang2018dynamic} and traffic management~\cite{gong2024real}, to micro-level tasks including travel behavior modeling ~\cite{li2023accurate}, trajectory imputation~\cite{yuan2024nuhuo,chen2023teri}, trajectory representation ~\cite{zhou2024red}, and next location recommendation~\cite{cai2024have}.
Despite the critical importance of urban trajectory data, the sensitivity of such data and associated privacy concerns~\cite{kapp2023generative,hu2024real} make the large-scale publication of \textit{fine-grained} trajectory data highly challenging.
In recent years, trajectory synthesis has emerged as an important research direction, aiming to generate synthetic trajectory data that closely align with real-world distributions, which can ensure data usability while minimizing the risk of privacy leakage. 

Existing trajectory synthesizing and generation studies can be divided into traditional non-generative methods and learning-based generative methods. 
Rule-based non-generative synthesizing solutions~\cite{nergiz2008towards} generate real trajectories by noise perturbation and multi-source data mixing. While these approaches offer some degree of privacy protection, they often distort the underlying data distribution, thereby compromising the utility of the generated data.
Existing learning-based generative trajectory synthesizing solutions can be further divided into autoregressive-based~\cite{chen2021trajvae,huang2019variational} and non-autoregressive-based~\cite{liu2018trajgans,jiang2023continuous}. 
Autoregressive-based methods fit well with discrete data, but suffer from constrained
sampling and cumulative errors, making them less suitable for modeling continuous trajectory structures. 
Non-autoregressive-based methods utilize generative models (e.g., VAE~\cite{kingma2013auto}, GAN~\cite{goodfellow2014gan}) to generate the entire trajectory holistically~\cite{rao2025seed},
which better suits the generation of fine-grained, continuous data.
However, these models suffer from low fidelity and reduced spatial validity, particularly in fine-grained trajectory synthesis tasks.

Recently, as a type of non-autoregressive generative method, diffusion models~\cite{ho2020denoising,dhariwal2021diffusion_beat_gan} have gained attention in trajectory synthesis due to their strong generative capacity and robustness. Some studies~\cite{wei2024diff,shi2024graph} apply diffusion models to generate road-segment-level paths or sequences, which enhances the realism of the synthesized trajectories but lacks fine spatial granularity. Others~\cite{zhu2023difftraj,zhu2024controltraj} directly generate fine-grained sequences of GPS points. However, these approaches either suffer from low spatial consistency and weak alignment with the road network~\cite{zhu2023difftraj} due to the complexity of real-world urban mobility, or are impractical to scale because of the high cost of acquiring road-segment annotations for each trajectory~\cite{zhu2024controltraj}.
For example, ControlTraj~\cite{zhu2024controltraj} introduces a controllable diffusion model conditioned on road-segment sequences during both training and inference, but the reliance on such fine-grained supervision signals, which are prohibitively costly to obtain, limits its scalability in practice.
In summary, although diffusion models are capable of modeling complex data distributions,
it remains non-trivial to apply diffusion models to fine-grained trajectory generation because of the following two key challenges:
(i) How to generate fine-grained trajectories that are both realistic and plausibly aligned with road network constraints, ensuring spatial consistency and structural validity; and
(ii) How to achieve efficient and scalable fine-grained trajectory synthesis when modeling complex real-world distributions without relying on extensive supervision.

To address the above challenges, in this paper, 
we leverage the opportunity that \textit{trajectories are naturally hierarchical in the spatial domain}~\cite{bassolas2019hierarchical}. 
By decomposing trajectories into coarse-grained road segments and fine-grained GPS sequences, 
we propose a \underline{\textbf{Ca}}scaded hyb\underline{\textbf{r}}id \underline{\textbf{diff}}usion model-based trajectory synthesizing framework (termed \textbf{Cardiff}) to generate fine-grained and realistic trajectories. 
Particularly, 
(i) To enhance spatial realism and ensure road-network validity, our cascaded diffusion framework first generates coarse-grained road segment sequences. The segment-level synthetic results then serve as structural conditions to guide fine-grained GPS trajectory generation through a noise-augmented cross-attention mechanism, progressively refining spatial details while preserving topological consistency.
(ii) To enable efficient and scalable fine-grained synthesis, we introduce a segment-level discrete-to-continuous latent autoencoder that compresses road segment sequences into compact latent representations. By operating the diffusion process in this low-dimensional latent space, we reduce computational complexity and improve generation efficiency.
In summary, the key contributions of our work are as follows.
\begin{itemize}[leftmargin=*]
    \item To the best of our knowledge, we are the first to integrate the spatial hierarchy into a cascaded diffusion framework for scalable and fine-grained trajectory generation. Through the coarse-to-fine cascaded principle, we first generate segment-level trajectories, thus ensuring the validity of the road network and enhancing the fine-grained diffusion-based generation through first-stage conditioning.
    \item Based on the coarse-to-fine cascaded diffusion paradigm, in the \textit{Cardiff} framework, we first design a segment-level latent diffusion module that compresses road segment sequences with varying lengths into compact latent representations, enabling efficient segment-level denoising in a low-dimensional space. 
    Then, in the fine-grained GPS level, conditioning on the coarse-grained generation results, a conditional continuous diffusion model is utilized to synthesize high-resolution GPS trajectories with improved spatial consistency. 
    \item We conduct extensive experiments on three large real-world trajectory datasets. Experimental results show that the proposed \textit{Cardiff} outperforms state-of-the-art baselines in various metrics that measure the realism of the synthetic trajectories. Moreover, we also show that the cascaded hybrid structure can effectively provide privacy guarantees while maintaining the downstream utilities. 
    
\end{itemize}

%% file: Sec2-Problem.tex
\section{Preliminary and Formulation}

\subsection{Preliminaries}

\begin{definition}[\textsc{\textbf{Trajectory}}] 
A trajectory $\tau$ is a sequence of continuous GPS points, $\tau = \{l_0, l_1, \cdots, l_i, \cdots, l_L\}$, where $l_i = \{lat_i, lon_i\}$ represents a GPS point with latitude and longitude. $L = |\tau|$ represents the length of trajectory $\tau$.
\end{definition}

\begin{definition}[\textsc{\textbf{Road Network}}] The road network $G = (V, E) $ is a direct graph, where $V$ means represents the road segments in road networks. $E$ is the edge set that represents the transition direction and topology between different road segments.
\end{definition}

\begin{definition}[\textsc{\textbf{Road Segment-level Trajectory}}] 
The road segment trajectory $\tau_r = \{s_1, s_2, \cdots, s_i, s_n\}$ means the road segment sequence of one GPS trajectory.  $s_i$ is the ID of the road segment $i$ in the road network $G$. The features of road segments consist of the attributes (e.g., length, types, speed) and geometry features (sequence of GPS coordinates).

\end{definition}

\begin{figure}[h] 
\centering 
\includegraphics[width=0.43\textwidth]{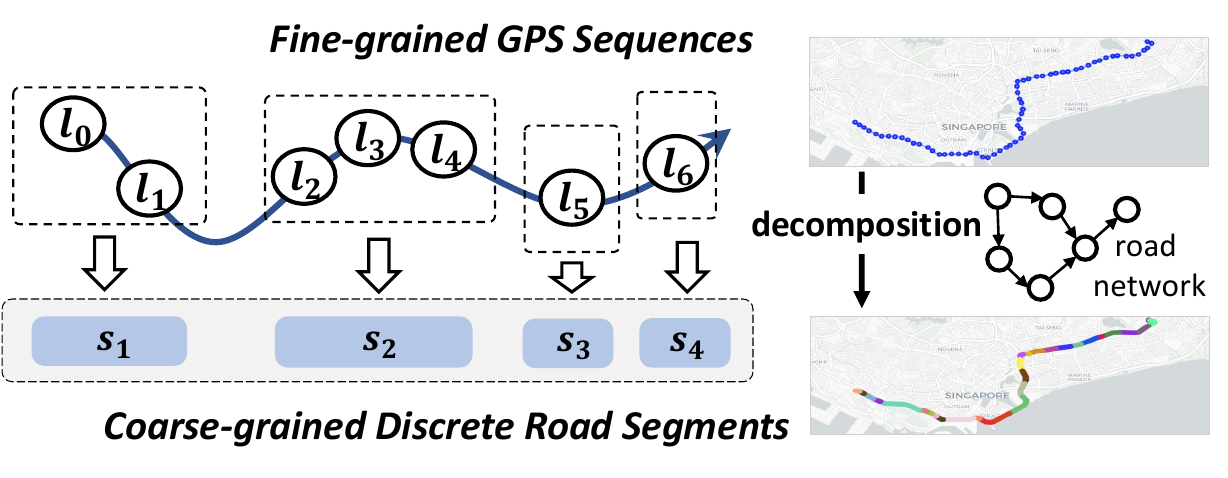} 
	\caption{The spatial hierarchy of trajectories. \label{fig:decomposition}}
\end{figure}

\begin{definition}[\textsc{\textbf{Trajectory Hierarchical Decomposition}}] 
Inspired by the analysis of the hierarchical organization of urban mobility~\cite{wikipedia_transportation_forecasting,bassolas2019hierarchical}, as shown in Fig.~\ref{fig:decomposition}, 
we decompose the trajectory instance into coarse-grained road segments sequence $\tau_r$ and fine-grained GPS sequence $\tau$. 
The key intuition of hierarchical decomposition is that human trajectories contain different levels of physical information. The coarse-grained road segment sequence maintains the transition patterns and traffic distribution among different road network segments, while the fine-grained GPS sequence can capture specific location information (e.g., home and working places) and microscopic mobility features. 

\end{definition}

\subsection{Diffusion Probabilistic Model}
\label{sec-sub-diffusion-preliminaries}
We first present a brief overview of the concepts and formulation of the denoising diffusion probabilistic model~(DDPM)~\cite{ho2020denoising} as well as the cascaded diffusion structure~\cite{ho2022cascaded}. 
The denoising diffusion probabilistic model consists of a forward diffusion process and a reverse denoising process. 

\textbf{Forward diffusion process:} 
During the forward process, Gaussian noise is gradually added into the input real sample $x_0$ step by step, until a Gaussian noise distribution $x_T$ in $T$ steps. 
The step sizes are controlled by a variance schedule $\{\beta_t \in (0, 1)\}_{t=1}^t$, which controls the noise injection ratio at each step.
Specifically, in each step $t$, the forward diffusion process, which is defined as: 
\begin{equation} \label{equ:Markov-forward}
    q(\mathbf{x}_t \vert \mathbf{x}_{t-1}) = \mathcal{N}(\mathbf{x}_t; \sqrt{1 - \beta_t} \mathbf{x}_{t-1}, \beta_t\mathbf{I}),\quad t=1,2,\cdots, T
\end{equation}
Based on the Markov chain of the forward diffusion process, we have $q(\mathbf{x}_{1:T} \vert \mathbf{x}_0) = \prod^T_{t=1} q(\mathbf{x}_t \vert \mathbf{x}_{t-1})$. Through reparameterization trick~\cite{ho2020denoising} and let $\alpha_t = 1 - \beta_t$ and $\bar{\alpha}_t = \prod_{i=1}^T \alpha_i$,  at any arbitrary time step $t$ the $\mathbf{x}_t$ can be sampled as: 
\begin{equation}\label{equ:forward}
    \mathbf{x}_t  = \sqrt{\bar{\alpha}_t}\mathbf{x}_0 + \sqrt{1 - \bar{\alpha}_t}\epsilon_t
\end{equation}
where $\epsilon_t \sim \mathcal{N}(0,1)$.  Equation~\ref{equ:forward} allows that $\mathbf{x}_t$ can be directly calculated given any arbitrary time step $t$ and the initial input $\mathbf{x}_0$.  

\textbf{Reverse denoising process:} 
In the reverse denoising process, the diffusion model aims to learn a denoising network to recreate the real data samples through $p(\mathbf{x}_{t-1} \vert \mathbf{x}_t)$ from a Gaussian noise input, $\mathbf{x}_T \sim \mathcal{N}(\mathbf{0}, \mathbf{I})$ step by step, until reconstructing $x_0$. 
The reverse process that sample $\mathbf{x}_{t-1}$ from $\mathbf{x}_t$ can be represented as: 
\begin{equation}
p_\theta(\mathbf{x}_{t-1} \vert \mathbf{x}_t) = \mathcal{N}(\mathbf{x}_{t-1}; \boldsymbol{\mu}_\theta(\mathbf{x}_t, t), \boldsymbol{\Sigma}_\theta(\mathbf{x}_t, t))
\end{equation}
According to DDPM~\cite{ho2020denoising}, the variance $\boldsymbol{\Sigma}_\theta(\mathbf{x}_t, t)$ is set to $\beta_t \mathbf{I}$ as untrained time dependent constants.
The mean $\boldsymbol{\mu}_\theta(\mathbf{x}_t, t) $ is estimated by a neural network parameterized by $\theta$, denoted as 
\begin{equation}
\label{equ:denoising-function}
   \boldsymbol{\mu}_\theta(\mathbf{x}_t, t) = \frac{1}{\sqrt{\alpha_t}} \Big( \mathbf{x}_t - \frac{\beta_t}{\sqrt{1 - \bar{\alpha}_t}} \mathbf{\epsilon}_\theta(\mathbf{x}_t, t) \Big)  
\end{equation}
Therefore, estimating the reverse transition distribution $p_\theta(\mathbf{x}_{t-1} \vert \mathbf{x}_t)$ is reparameterized to predict $\mathbf{\epsilon}$ given 
input $\mathbf{x}_t$ at time step $t$~\cite{nichol2021improved}.

The training of the noised estimated network (denoising network ) $\mathbf{\epsilon}_\theta(x_t, t)$ is based on the variational lower bound (VLB)~\cite{kingma2013auto}, which is defined as $\mathcal{L}_\text{VLB} = \mathcal{L}_T + \mathcal{L}_{T-1} + \dots + \mathcal{L}_0$. 
For each step
$t$, denoising score matching loss $\mathcal{L}_t$ is a distance between
two Gaussian distributions (i.e., posterior $q(\mathbf{x}_t \vert \mathbf{x}_{t+1}, \mathbf{x}_0)$ and predicted reverse distribution $p_\theta(\mathbf{x}_t \vert\mathbf{x}_{t+1})$), which can be rewritten in terms
of noise predictor $\mathbf{\epsilon}_\theta$ as:
\begin{equation}
\begin{aligned}
\mathcal{L}_t 
&= D_\text{KL}(q(\mathbf{x}_t \vert \mathbf{x}_{t+1}, \mathbf{x}_0) \parallel p_\theta(\mathbf{x}_t \vert\mathbf{x}_{t+1})) \\
&= \mathbb{E}_{\mathbf{x}_0, \mathbf{\epsilon}} \Big[\frac{ \beta_t^2 }{2 \alpha_t (1 - \bar{\alpha}_t) \| \boldsymbol{\Sigma}_\theta \|^2_2} \|\mathbf{\epsilon}_t - \mathbf{\epsilon}_\theta(\mathbf{x}_t, t)\|^2 \Big] \\
\end{aligned}
\end{equation}
Empirically, based on DDPM~\cite{ho2020denoising}, the $\mathcal{L}_t $ can be further reduced with a simplified objective that ignores the weighting term:
\begin{equation}
\label{equ:traing_loss}
    \mathcal{L}_t^\text{simple} = \mathbb{E}_{\mathbf{x}_0, \mathbf{\epsilon}_t} \Big[\|\mathbf{\epsilon}_t - \mathbf{\epsilon}_\theta(\sqrt{\bar{\alpha}_t}\mathbf{x}_0 + \sqrt{1 - \bar{\alpha}_t}\mathbf{\epsilon}_t, t)\|^2 \Big]
\end{equation}

\textbf{Cascaded Diffusion Model:}
Cascaded diffusion models~\cite{ho2022cascaded,saharia2022photorealistic,ramesh2022hierarchical} consists
of multiple diffusion models that generate images of increasing resolution, beginning with a
standard diffusion model at the lowest resolution, followed by one or more super-resolution
diffusion models that refine higher resolution details to improve the overall quality.
Existing cascaded diffusion models, such as the cascaded DDPM~\cite{ho2022cascaded} and Imagen~\cite{saharia2022photorealistic} work with isomorphic data of varying resolutions at different levels. However, in our scenario, \textit{the unique aspect is that the hierarchical nature of trajectory data is partitioned into discrete road segments and continuous GPS points, which poses additional challenges.} More details of our cascaded hybrid design are illustrated in Section~\ref{sec:model}.

\subsection{Problem Formulation}
In this paper, the overall goal is to generate high-fidelity fine-grained urban mobility trajectories, which should not only share a high similarity with the real-world trajectory distribution to retain the utility for downstream spatial-temporal applications but also provide flexible privacy guarantees for large-scale data publishing purposes.
Taking the hierarchical structure of urban mobility as an opportunity, we decompose the trajectory data into coarse-grained segment-level trajectories and fine-grained GPS-level sequences. 
The trajectory generation can be performed in a multi-granularity manner progressively.

Formally, the trajectory generation problem studied in this paper can be defined as:
Given a set of historical trajectories $\mathcal{T} = \{\tau_1, \tau_2, \cdots, \tau_N\}$, the objective of coarse-to-fine trajectory generation is to learn a cascaded generative model $\mathcal{F}_{cas}$. For each sample, $\mathcal{F}_{cas}$ first generates the coarse-grained road segment trajectory $\hat{\tau}_r$ through the segment-level generation function $\mathcal{F}_r$. Then, conditioned by $\hat{\tau}_r$, a second-stage GPS-level generation function $\mathcal{F}_g$ is used to synthesize the fine-grained GPS trajectory $\hat{\tau}$.

%% file: Sec3-method.tex
\section{Model Design}\label{sec:model}

\subsection{System Overview of Cardiff}
To generate high-fidelity and fine-grained trajectory data, 
we leverage the spatial hierarchy property of urban mobilities and perform cascaded diffusion processes to generate road-segment level trajectories and fine-grained GPS trajectories in a coarse-to-fine manner. 

\begin{figure*}[h] 
\centering 
\includegraphics[width=0.72\textwidth]{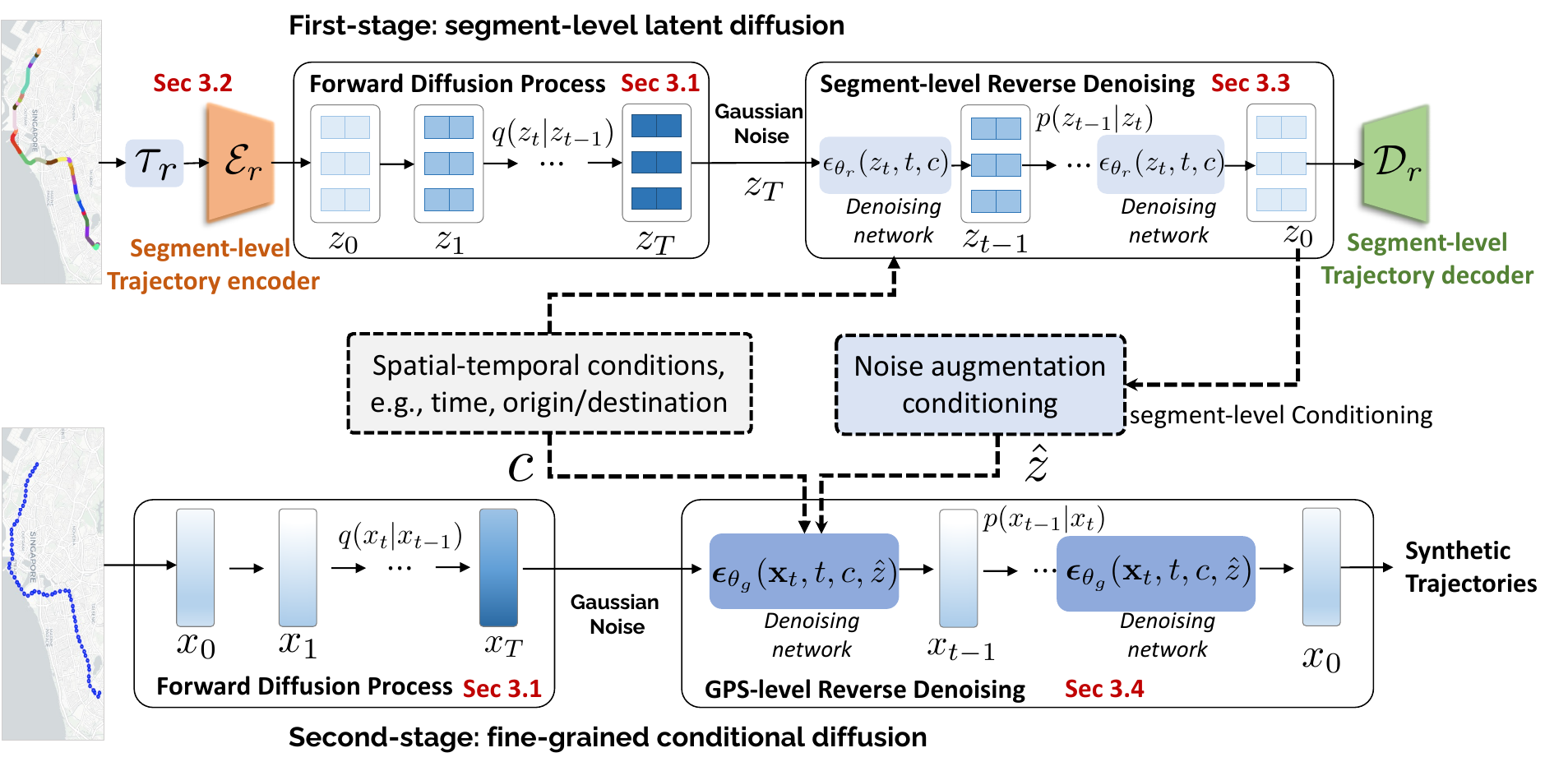} 
	\caption{The framework of \textsc{\textbf{Cardiff}}. \label{fig:framework}}
\end{figure*}

Fig.~\ref{fig:framework} illustrates our cascaded hybrid diffusion-based trajectory generation framework, named \textit{Cardiff}.
\textit{Firstly}, \textit{Cardiff} encodes the discrete segment-level trajectories into the continuous latent space with an autoregressive trajectory autoencoder (detailed in Sec.~\ref{sec:autoencoder}). 
\textit{Secondly}, a segment-level latent diffusion (Sec.~\ref{sec:first-level-Cascaded-design}) is utilized to denoise and generate the coarse-grained segment-level trajectory embeddings.
\textit{Lastly}, a conditional fine-grained diffusion module (Sec.~\ref{sec:second-level-Cascaded-design}) is proposed to generate the high-fidelity GPS-level trajectories progressively conditioned on the segment-level latent. 

\subsubsection{Motivation for Utilizing Cascaded Diffusion}
The cascaded diffusion design naturally aligns with the hierarchical structure of urban trajectory data. 
Specifically, it offers the following advantages:
(i) During training, the coarse-to-fine structure allows the model to effectively capture the multi-spatial granularity of trajectory data. The segment-level modeling focuses on learning the structural backbone and semantic patterns, while the fine-grained denoising stage concentrates on detail refinement. 
(ii) In the sampling, the cascaded framework enables progressive trajectory synthesis from pure noise without external supervision (e.g., road network), achieving both road network validity and fine-grained spatial fidelity.

\subsubsection{Overview of Parallelized Forward Processes}
As discussed in section~\ref{sec-sub-diffusion-preliminaries}, the forward diffusion process in diffusion probabilistic models is modeled as a Markov process $q(\mathbf{x}_t \vert \mathbf{x}_{t-1})$, which 
gradually add gaussian noise into the input sample $x_0$ step by step, until a Gaussian noise distribution $x_T$ in $T$ steps. 
In our cascaded diffusion framework,
as illustrated in Fig.~\ref{fig:framework}, 
we decompose the trajectory data into the coarse-grained road-segment level and fine-grained continuous GPS level. 
The forward processes of the two-level diffusion are conducted in parallel. 
According to the Markov chain formulation (Equation~\ref{equ:Markov-forward})  and the forward diffusion process (Equation~\ref{equ:forward}),
in the coarse-grained level, 
the forward diffusion is performed on the latents of the road segment sequence and is modeled as $q(\mathbf{z}_t \vert \mathbf{z}_{t-1}) = \mathcal{N}(\mathbf{z}_t; \sqrt{1 - \beta_t} \mathbf{x}_{t-1}, \beta_t\mathbf{I})$. 
Given an arbitrary diffusion step $t$, 
we have 
\begin{equation}\label{equ:traj-first-forward}
        \mathbf{z}_t  = \sqrt{\bar{\alpha}_t}\mathbf{z}_0 + \sqrt{1 - \bar{\alpha}_t}\epsilon_{t_r}
\end{equation}
where $\mathbf{z}_0$ is the clean latent representation of the input segment-level trajectory. $\bar{\alpha}_t$ is the constant hyperparameter to control the noise injection degree for sampling. 
$\epsilon_{t_r}$ is a noise sampled from the Gaussian distribution.

At the fine-grained GPS-level forward diffusion, 
we define the clean input $x_o$ as the fine-grained trajectory sample $\tau$.
We can then obtain the $x_t$ at any given diffusion step $t$ as
$ \mathbf{x}_t  = \sqrt{\bar{\alpha}_t}\mathbf{x}_0 + \sqrt{1 - \bar{\alpha}_t}\epsilon_{t_g} $, 
where $\mathbf{x}_0$ is the input fine-grained trajectory sample. 
$\epsilon_{t_g}$ is the sampled GPS-level Gaussian random noise.
The forward diffusion process produces a sequence of noisy samples, which can be used as training data for denoising network training.

\subsubsection{Overview of Cascaded Reverse Denoising}
The reverse denoising process $p_\theta(\mathbf{x}_{t-1} \vert \mathbf{x}_t)$ gradually anneal the noise from a gaussian distribution $x_T$ to a data distribution $p(x_0)$ by learning the noise prediction function $\boldsymbol{\mu}_\theta(\mathbf{x}_t, t)$. 
Suppose $z_0$ is the latent of the discrete road-segment trajectory $\tau_r$ and $x_0$ is the corresponding fine-grained GPS-level trajectory $\tau$.
As shown in Fig.~\ref{fig:framework}, our cascaded trajectory denoising pipeline consists of two denoising stages: (i) a coarse-grained road-segment denoising model to recover $z_0$, (ii) a continuous GPS denoising model to obtain the final fine-grained trajectory sample $x_0$. Additionally, a noise augmentation conditioning mechanism is proposed to bridge the coarse-to-fine denoising processes.

\subsection{Encoding Discrete Segments into Latents}
\label{sec:autoencoder}

Latent diffusion models~\cite{rombach2022high} have been widely applied to image generation tasks, where VAEs~\cite{kingma2013auto} are used to compress the dimensionality of training samples to reduce the complexity of the diffusion process. 
However, for generating discrete road segment sequences, existing research lacks an exploration of the trajectory encoding tailored for generative tasks. 
To bridge the gap, we observe that \textit{modeling discrete road segment sequences share significant similarities with natural language processing, where each road segment can be regarded as a language token and a trajectory composed of these segments can be analogized to a semantically meaningful sentence.}

Therefore, inspired by latent encoding processes in natural language generation~\cite{lewis2019bart,zhang2023planner,lovelace2023latent}, 
we developed an autoregressive trajectory autoencoder, which is similar to the Bidirectional and Auto-Regressive Transformers (BART) model~\cite{lewis2019bart} and consists of 
(i) a Bert-style encoder $\mathcal{E}$ with multiple transformer-encode layers to transform the discrete road segment sequences into continuous latent feature spaces, and
(2) a GPT2-like auto-regressive decoder $\mathcal{D}$ with multiple transformer-decode layers to reconstruct original segment sequences from latent. 
This approach delegates the generation of high-frequency details to the autoencoder and reduces the complexity of the diffusion learning process. 

\begin{figure}[h] 
\centering 
\includegraphics[width=0.48\textwidth]{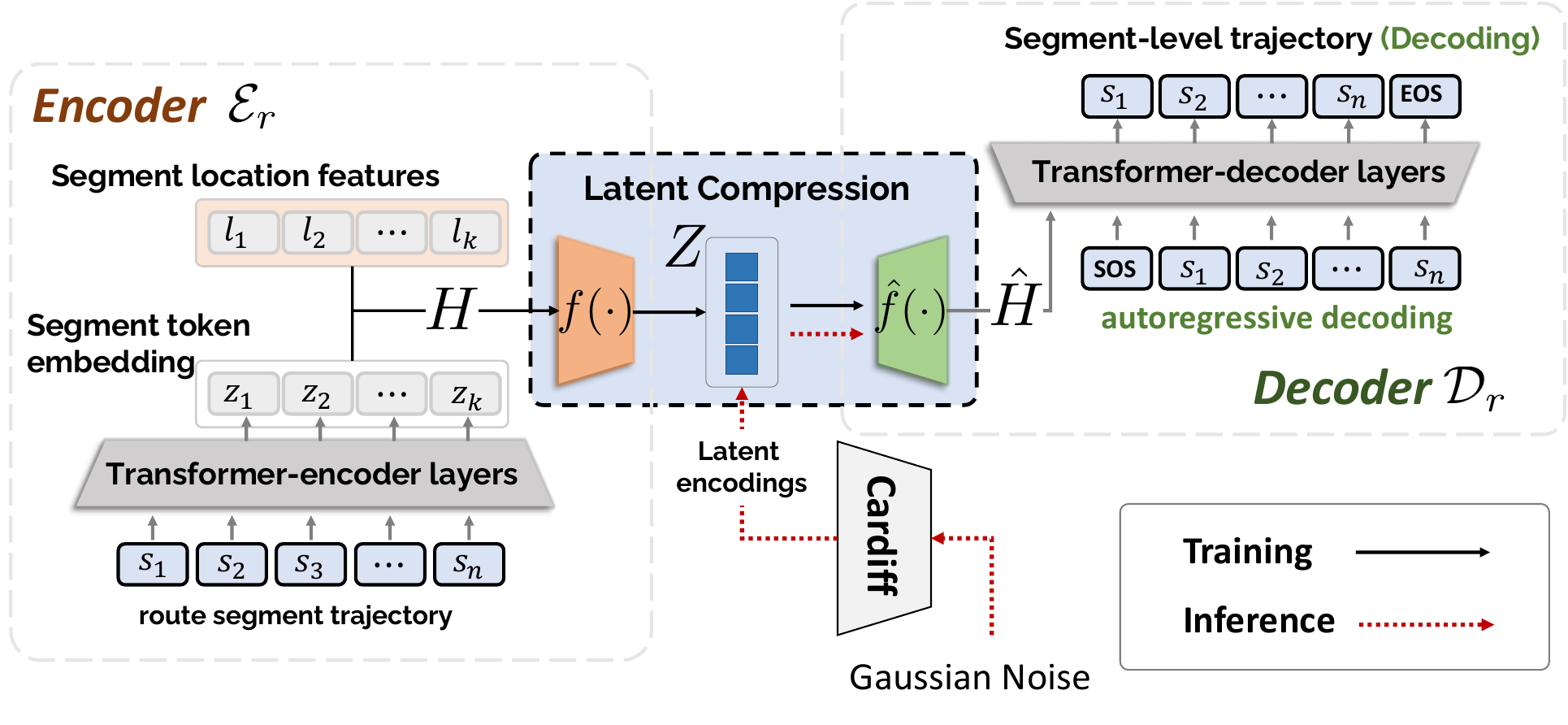} 
	\caption{Illustration of trajectory autoencoder.\label{fig:ae}} 
\end{figure}

\textbf{Segment-level latent encoding and compression:}
As shown in Fig.~\ref{fig:ae}, our trajectory encoder consists of two components: a stack of Transformer encoder layers for segment-level trajectory encoding, and a cross-attention-based latent compression module inspired by the perceiver architecture~\cite{jaegle2021perceiver}.
Given an input trajectory represented as a variable-length sequence of segments \(\tau_r = \{s_1, s_2, \dots, s_n\}\), we first obtain token-level representations using the stacked Transformer encoder:
\begin{equation}
\mathbf{H} = \mathrm{TransformerEncoder}(s_1, s_2, \dots, s_n)
\end{equation}
To incorporate spatial information, we embed the center-point coordinates of each segment and integrate them into the segment-level token embeddings of \(\mathbf{H}\) before feeding them into the compression module.
Then, to obtain a compact representation suitable for diffusion modeling, we introduce latent compression layers $f(\cdot)$ into the encoder. Specifically, a set of learnable latent vectors \(\mathbf{Z}^{(0)} \in \mathbb{R}^{L \times d}\) is attended to the encoder outputs \(\mathbf{H}\) via cross-attention:
\begin{equation}
\mathbf{Z} = f(\mathbf{Z}^{(0)}, H) = \mathbf{Z}^{(0)} + \mathrm{CrossAttn}\bigg(Q(\mathbf{Z}^{(0)}), K(\mathbf{H}), V(\mathbf{H})\bigg)
\end{equation}
The resulting fixed-length, compact latent representation \(\mathbf{Z} \in \mathbb{R}^{L \times d}\) serves as the input to the segment-level latent diffusion model.

\textbf{Autoregressive decoding:}
During decoding, we first map $\mathbf{Z}$ back into an expanded latent sequence $\hat{\mathbf{H}}$ using a learnable cross-attention-based decompression network $\hat{f}(\mathbf{Z})$, which serves as the inverse transformation of the compression module used in the encoder. 
After obtaining $\hat{\mathbf{H}}$, we then feed it into the autoregressive decoder with multiple transformer decoder layers to reconstruct the segment-level trajectory, token by token through:
$
\hat{s}_l = \mathrm{TransformerDecoder}(\hat{s}_{<l}, \hat{\mathbf{H}})
$.

For model training, the latent trajectory autoencoder model is trained using teacher forcing to minimize the negative log-likelihood between the generated and ground-truth segment trajectories:
\begin{equation}
\mathcal{L}_{\mathrm{rec}} = -\sum_{l=1}^{L} \log P(\hat{s}_l = s_l \mid \hat{s}_{<l}, \hat{\mathbf{H}}),
\end{equation}
where $L$ is the length of the road segment-level trajectory $\tau_r$. $s_l$ means the groundtruth of $l$-th segment id and $\hat{s}_l$ is the predicted one.
During inference, 
as shown at the bottom of Fig.~\ref{fig:ae}, 
we first sample a compact latent vector $\hat{\mathbf{Z}}$ from the latent diffusion model of \textit{Cardiff}. 
The sampled latent is then decompressed and decoded. 
The decoding is performed token-by-token in an autoregressive manner using beam search, ultimately yielding a complete segment-level trajectory sequence.

\subsection{Coarse-grained Segment-level Latent Denoising}
\label{sec:first-level-Cascaded-design}

In the first stage, the denoising process operates in the segment-level latent space. 
At denoising step $t$, the road-segment level denoising function $p_\theta (z_{t-1} \vert z_{t}, c)$ is denoted as 
\begin{equation}
    p_\theta (z_{t-1} \vert z_{t}, c) = \mathcal{N}(z_{t-1}, \boldsymbol{\mu}_{\theta_r}(\mathbf{z}_t, t \vert c, \sigma^2\mathbf{I})) \quad t=T,\cdots,2, 1
\end{equation}
where ${\theta_r}$ includes the parameters of the denoising network at the road segment level.
$c$ is the external conditional information, which consists of the departure time, origin, destination, and some statistics (e.g., lengths, duration) of trajectories. 
$\boldsymbol{\mu}_{\theta_r}(\mathbf{z}_t, t \vert c, \sigma^2\mathbf{I})$ is Gaussian noise distribution at time step $t$.
According to Equation~\ref{equ:denoising-function}, through reparameterize,
$\boldsymbol{\mu}_{\theta_r}(\mathbf{z}_t, t \vert c, \sigma^2\mathbf{I})$ can be obtain by 
\begin{equation}
    \boldsymbol{\mu}_{\theta_r}(\mathbf{z}_t, t \vert c, \sigma^2\mathbf{I}) 
    = \frac{1}{\sqrt{\alpha_t}} \Big( \mathbf{z}_t - \frac{\beta_t}{\sqrt{1 - \bar{\alpha}_t}} \boldsymbol{\epsilon}_{\theta_r}(\mathbf{z}_t, t \vert c) \Big)
\end{equation}
where $\boldsymbol{\epsilon}_{\theta_r}(\mathbf{z}_t, t \vert c)$ is noise estimated network for the segment-level denoising process. 
$\mathbf{z}_t$ is the noised latent segment-level representations at diffusion step $t$.
$c$ is the spatial-temporal contextual conditions. 
So the learning goal of road segment-level denoising is to learn the noise estimated network $\boldsymbol{\epsilon}_{\theta_r}(\mathbf{z}_t, t \vert c)$.

\begin{figure}[h] 
\centering 
\includegraphics[width=0.35\textwidth]{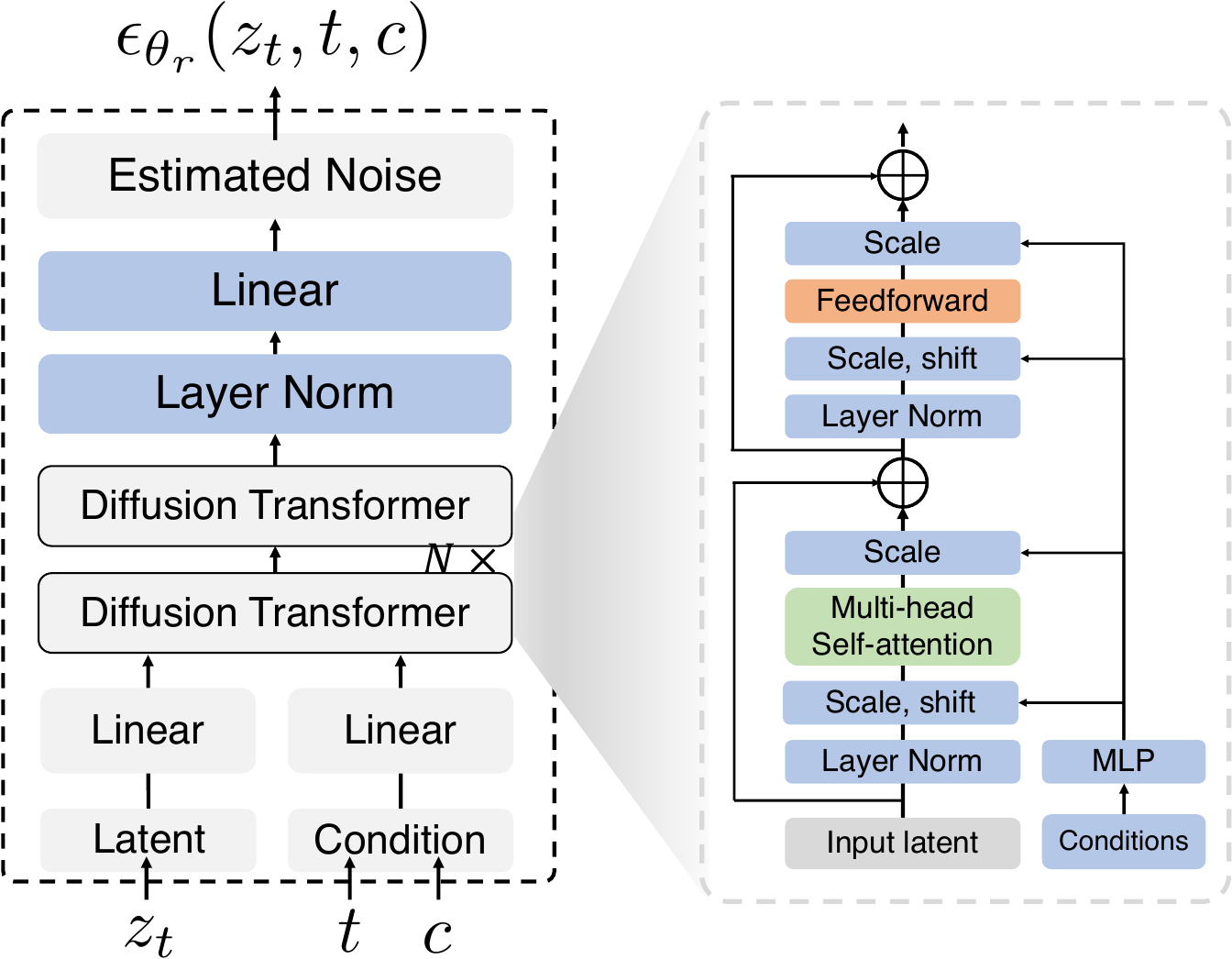} 
	\caption{Structure of segment-level denoising network.\label{fig:first-level}}
\end{figure}

\subsubsection{Segment-level Latent Denoising Network}
As illustrated in Fig.~\ref{fig:first-level}, in the coarse-grained latent denoising process, the proposed segment-level latent denoising network is built on diffusion transformer (DiT) blocks~\cite{peebles2023scalable} to model the spatial representations of discrete road-segment sequences and predict the noise. 
The input of the segment-level denoising network includes the noisy latent  $z_t$, the diffusion step $t$, and the attribute features $c$ of each trajectory. 
$t$ and $c$ are utilized as conditional signals and are integrated into the DiT layers via an adaptive layer normalization mechanism~\cite{peebles2023scalable} to enable effective feature fusion and conditional control. 
After the final diffusion transformer block, we employ a standard linear decoder to obtain the predicted noise $\boldsymbol{\epsilon}_{\theta_r}(\mathbf{z}_t, t \vert c)$ at each step $t$.

\subsubsection{Training Objectives:}
The segment-level denoising loss is formulated as the difference between the ground-truth noise and the predicted noise by the noise estimated denoising network $\boldsymbol{\epsilon}_{\theta_r}(\mathbf{z}_t, t \vert c)$, which is denoted as:
\begin{equation}
\mathcal{L}^\text{segment} = \mathbb{E}_{t, \mathbf{z}_t, \mathbf{\epsilon}} \left[\|\epsilon_t - \mathbf{\epsilon}_{\theta_r}(z_{t}, t, c)\|^2\right],
\end{equation}
where ${\theta_r}$ denotes the parameters of road segment-level denoising network, and $z_{t_r}$ is the noised latent at step $t$.
To ensure the generated segment sequences follow the real-world road topology, 
we introduce a spatial validity loss:
\begin{equation}
\mathcal{L}_{\text{phy}} = \frac{1}{L - 1} \sum_{\ell=1}^{L - 1} \left(1 - A_{s_\ell, s_{\ell + 1}} \right),
\end{equation}
where $A$ is the binary adjacency matrix of the road network, and $A_{s_\ell, s_{\ell+1}} = 1$ if two consecutive segments are physically connected. 
Unlike existing works~\cite{rao2025seed,wei2024diff} that directly apply the spatial validity as generation constraints or loss terms throughout all denoising steps, we adopt an adaptive physics-informed strategy~\cite{xu2023interdiff,yuan2023physdiff}, applying spatial constraints only in the later stages of the denoising process, where semantic structure becomes more prominent. In contrast, early denoising steps operate in a high-noise regime and offer limited structural information.
The overall training objective for the segment-level denoising network is:
\begin{equation}
\label{equ:first-level-loss}
\mathcal{L}_{r} = \mathcal{L}^\text{segment} + \lambda_{p} \mathcal{L}_{\text{phy}},
\end{equation}
where $\lambda_{p}$ is a balancing hyperparameter controlling the strength of the spatial validity constraint.

\subsection{Fine-grained GPS-level Conditional Denoising}
\label{sec:second-level-Cascaded-design}
In our \textit{Cardiff} framework, as illustrated in Figure~\ref{fig:second-level}(a), after obtaining the coarse-grained denoising latent $z_0$ through segment-level denoising, 
the second-stage GPS-level denoising process generates high-resolution trajectories conditioned on noise augmented segment-level latent $\hat{z}$ and trajectory-level attributes $c$. 
Let $x_0$ as the GPS-level fine-grained trajectory sample $\tau$, the GPS-level denoising network $\boldsymbol{\epsilon}_{\theta_g}(\mathbf{x}_t, t, c, \hat{z})$ is trained to estimate the injected noise at each diffusion step $t$, which enables step-by-step reverse denoising to generate the final fine-grained GPS sequence.

\subsubsection{Fine-grained Conditional Denoising Network}
\begin{figure}[h] 
\centering 
\includegraphics[width=0.47\textwidth]{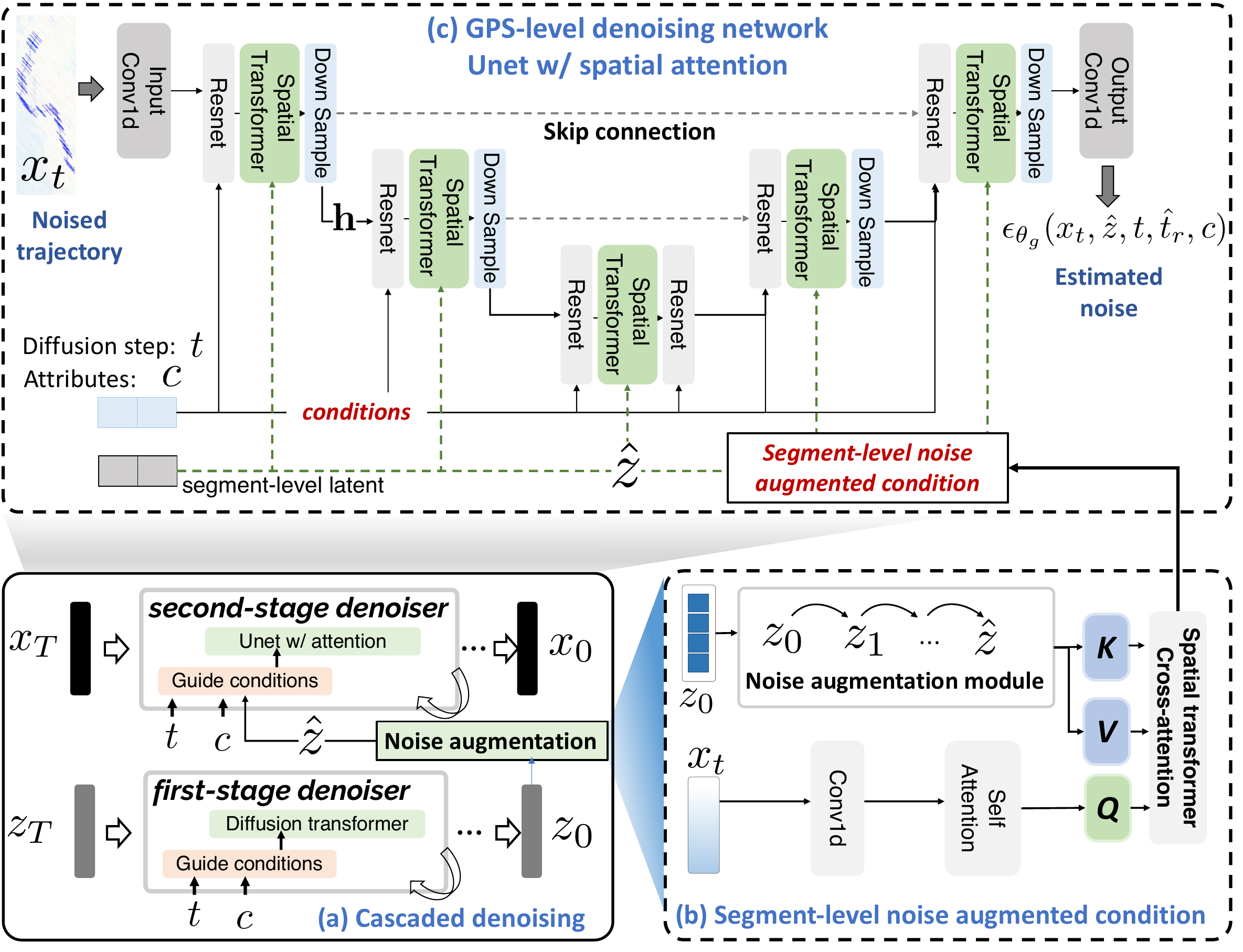} 
	\caption{(a) Cascaded denoising structure (bottom left), (b) Noise augmentation conditioning (bottom right), and (c) fine-grained GPS-level conditional denoising network (top).\label{fig:second-level}} 
\end{figure}

As shown in Figure~\ref{fig:second-level}(c), the GPS-level denoising network $\boldsymbol{\epsilon}_{\theta_g}(\mathbf{x}_t, t, c, \hat{z})$ is constructed based on the U-Net architecture equipped with the spatial cross-attention conditioning mechanism. 
The input noisy trajectories $\mathbf{x}_t \in \mathbb{R}^{N \times 2}$, where latitude and longitude are treated as two separate feature channels over a sequence of length $N$, are first mapped into high-dimensional hidden representations $\mathbf{h}$ via a 1D convolutional embedding layer, and then processed by a stack of up/down sampling blocks.
Each sampling block consists of several ResNet layers~\cite{he2016deep} and is augmented with a spatial-transformer layer~\cite{rombach2022high}. 
In the cross-attention of spatial-transformer layers, the hidden vector $\mathbf{h}$ is set as the query, and the noise-augmented segment-level latent $\hat{z}$ is projected and used as key and value inputs (see Equation~\ref{equ:cross-attention-sdn}). 
In addition, we also incorporate basic trajectory-level statistics as conditions $c$, such as the start and end locations and the departure timestamp, which are fused into the ResNet layers to provide global semantic context during the denoising process.
The trajectory denoising Unet structure and attention mechanism allow the $\boldsymbol{\epsilon}_{\theta_g}$ 
to better capture the multi-granularity spatial properties of fine-grained trajectories and generate spatially consistent as well as realistic outputs.

\subsubsection{Noise Augmented Cross-attention Conditioning}

To improve the robustness of the second-stage generation, we introduce a noise augmentation mechanism applied to the clean segment-level latent $z_0$ before conducting the cross-attention in spatial transformer layers. 
As shown in Figure~\ref{fig:second-level}(b), we randomly sample an augmentation timestep $\hat{t}_r \sim \text{Uniform}(\{1, \dots, T\})$ and apply the closed-form forward Equation~\ref{equ:forward} to produce the corrupted augmentation:
$\hat{z} = \sqrt{\bar{\alpha}_{\hat{t}_r}} z_0 + \sqrt{1 - \bar{\alpha}_{\hat{t}_r}} \cdot \epsilon, \quad \epsilon \sim \mathcal{N}(0, \mathbf{I})$.
The augmented latent $\hat{z}$ is then projected into key and value matrices and used as the conditioning input in the spatial transformer layers. Cross-attention is performed between the GPS-level trajectory hidden representation $\mathbf{h}$ of $\mathbf{x}_t$,  and the coarse-level latent $\hat{z}$ according to:
\begin{equation}
\label{equ:cross-attention-sdn}
\mathrm{CrossAttn}(\mathbf{h}, \tilde{z}) = \mathrm{softmax}\left(\frac{Q(\mathbf{h}) K(\hat{z})^\top}{\sqrt{d}}\right) V(\hat{z})
\end{equation}
where $Q(\cdot)$, $K(\cdot)$, $V(\cdot)$ are linear projection layers. $d$ is the feature dimension of query $Q(\mathbf{h})$.
The noise augmentation and cross-attention-based conditioning mechanism enable the denoising network to better follow the segment-level condition information and improve the quality and robustness of fine-grained GPS-level trajectory generation.

\subsubsection{Conditional Training of Fine-grained Denoising Network}

During training, the corrupted latent $\hat{z}$ along with the corresponding noise level $\hat{t}_r$ is provided as an additional input to the GPS-level denoising network $\epsilon_{\theta_g}$. 
The conditional training objective is to minimize the loss between noise $\epsilon_t$ at each diffusion step and the estimated noise $\mathbf{\epsilon}_{\theta_g}$ conditioned on the coarse-level output $\hat{z}$ by:
\begin{equation}
\label{equ:fine-grained-loss}
\mathcal{L}_g = \mathbb{E}_{t, \hat{t}_r, \mathbf{x}_0, \hat{z}, \mathbf{\epsilon}_t} \left[\|\epsilon_t - \mathbf{\epsilon}_{\theta_g}(x_{t}, \hat{z}, t, \hat{t}_r, c)\|^2\right],
\end{equation}
Where $c$ denotes additional trajectory-level attributes. $\mathbf{\epsilon}_{\theta_g}$ is the GPS-level noise estimated network.

\subsection{Cascaded Training and Sampling}

In this subsection, we first describe the training process of our cascaded hybrid diffusion framework based on real-world trajectory data.
Then, we detail the sampling procedures of the conditional cascaded design, where trajectory generation starts from pure Gaussian noise and progressively refines the high-fidelity trajectories through the trained cascaded denoising networks.

\begin{algorithm}[ht]
\caption{Training with Cascaded Hybrid Diffusion}
\label{alg:training}
\KwIn{Segment-level trajectories $\tau_r$, GPS-level trajectories $\tau$, encoder $\mathcal{E}$, and trajectory condition $c$}
Initialize parameters of segment-level and GPS-level denoisers $\theta_r$, $\theta_g$\;
Repeat epoch training of two cascaded denoisers $\theta_r$, $\theta_g$\;
\For{each mini-batch $(\tau_r, \tau, c)$}
{
    \algcommentline{Parallelized forward diffusion}
    Obtain Segment-level latent $z_0$ through $\mathcal{E}(\tau_r)$\;
    Set GPS-level trajectory $\tau$ as $x_0$\;
    Sample Segment-level $t_r$ and noise $\epsilon_r \sim \mathcal{N}(0, I)$\;
    Sample GPS-level $t_g$ and noise $\epsilon_t \sim \mathcal{N}(0, I)$\;
    \algcommentlineComment{Obtain noised sample based on Equation~\ref{equ:forward}}
    Segment-level: $z_{t_r} = \sqrt{\bar{\alpha}_{t_r}} z_0 + \sqrt{1 - \bar{\alpha}_{t_r}} \epsilon$\;
    GPS-level: $x_{t_g} = \sqrt{\bar{\alpha}_{t_g}} x_0 + \sqrt{1 - \bar{\alpha}_{t_g}} \epsilon$\;
    \algcommentline{Cascaded conditional training}
    Compute $\mathcal{L}_r = \|\epsilon_{t_r} - \epsilon_{\theta_r}(z_{t_r}, {t_r}, c)\|^2$ via Equ.~\ref{equ:first-level-loss}\;
    sample augmented segment-level condition $\hat{t}_r$ and $\hat{z}$\;
    Compute $\mathcal{L}_g = \|\epsilon_{t_g} - \epsilon_{\theta_g}(x_{t_g}, {t_g}, c|\hat{z},\hat{t}_r)\|^2$ via Equ.~\ref{equ:fine-grained-loss}\;
    Update $\theta_r$, $\theta_g$\;
}
until converged\;
\end{algorithm}

\subsubsection{Cascaded Hybrid Training}
\label{sec:cascaded-hybrid-training}

As shown in Algorithm~\ref{alg:training}, our \textit{Cardiff} framework adopts a cascaded hybrid training paradigm, which optimizes two diffusion-based denoising networks in a coarse-to-fine manner.  
We begin by performing forward diffusion sampling in a parallelized way to obtain noisy segment-level and GPS-level training samples, denoted as $z_t$ and $x_t$.

In the first stage, the road segment-level latent denoising network $\mathbf{\epsilon}_{\theta_r}$ is trained to reconstruct coarse-grained discrete trajectories in the latent space, based on the input $z_{t_r}$, diffusion step $t_r$, and condition $c$.  
The second-stage GPS-level denoising network $\mathbf{\epsilon}_{\theta_g}$ is trained to generate fine-grained continuous trajectories $x_0$, guided by both the corresponding noisy GPS-level input $x_t$ and the noise augmented segment-level latent $\hat{z}$.
This sequential training strategy enables \textit{Cardiff} to effectively learn multi-scale spatiotemporal patterns.

\begin{algorithm}[h]
\caption{Conditional Cascaded Sampling}
\label{alg:cdm_inference}
\KwIn{Trajectory condition $c$, trained denoising networks $\epsilon_{\theta_r}$, $\epsilon_{\theta_g}$, 
and noise schedule $\beta_t$}
Sample $z_T \sim \mathcal{N}(0, I)$\;
\algcommentline{Coarse-grained denoising stage}
    \For{$t = T$ to $1$}{
        Sample initial noise noise $\epsilon \sim \mathcal{N}(0, I)$ if $t > 1$, else $\epsilon = 0$\;
        Compute denoised sample $ z_{t-1}$ based on reverse function:
        \[
        z_{t-1} = \frac{1}{\sqrt{\alpha_t}} \left( z_t - \frac{\beta_t}{\sqrt{1 - \bar{\alpha}_t}} \epsilon_{\theta_r}(z_t, t, c) \right) + \sigma_t \epsilon
        \]
    }
Obtain segment-level denoising sample $z_0$\;  

Sample initial noise $x_T \sim \mathcal{N}(0, I)$\;
\algcommentline{Conditional fine-grained denoising stage}
    \For{$t = T$ to $1$}{
        Sample noise $\epsilon \sim \mathcal{N}(0, I)$ if $t > 1$, else $\epsilon = 0$\;
        Compute the denoised sample based on the first-stage generated results $z_0$:
        \[
        x_{t-1} = \frac{1}{\sqrt{\alpha_t}} \left( x_t - \frac{\beta_t}{\sqrt{1 - \bar{\alpha}_t}} \epsilon_{\theta_g}(x_t, t, c, z_0) \right) + \sigma_t \epsilon
        \]
    }
\Return fine-grained synthetic trajectory $x_0$\;
\end{algorithm}

\subsubsection{Cascaded Conditional Sampling}

During inference, the \textit{Cardiff} framework adopts a \textit{cascaded conditional sampling} strategy to progressively generate high-fidelity trajectories from pure Gaussian noise.  
The full sampling procedure is detailed in Algorithm~\ref{alg:cdm_inference}.  
In the first stage, we initialize the latent of road segment-level trajectory from random noise $\mathbf{z}_T \sim \mathcal{N}(0, \mathbf{I})$,  
and perform reverse diffusion guided by the predicted noise $\epsilon_{\theta_r}(z_t, t, c)$ at each timestep $t$.  
Once the segment-level latent $z_0$ is generated, it serves as an auxiliary condition for the second-stage sampling and provides road network constraints to the fine-grained generation.

In the second stage, we initialize the GPS-level trajectory from pure noise $\mathbf{x}_T \sim \mathcal{N}(0, \mathbf{I})$,  
and perform conditional reverse diffusion, where the denoising network $\epsilon_{\theta_g}(x_t, t, c, z_0)$ predicts the noise at each step.  
During sampling, we directly use the generated segment-level latent $z_0$ as the condition for GPS-level denoising, without applying noise augmentation.
By leveraging both the global condition $c$ and the previously generated segment-level latent $z_0$, the fine-grained denoising process ensures that the final GPS-level trajectory $x_0$ remains globally consistent while capturing finer spatial-temporal details.
This hierarchical conditional generation scheme enables \textit{Cardiff} to maintain road network structural coherence across scales while achieving high-fidelity trajectory synthesis.

%% file: Sec4-evaluation.tex
\section{Evaluation}
In this section, we first introduce the experimental setup, including datasets, implementation details, baselines, and evaluation metrics.  
Then, we conduct extensive experiments to answer the following research questions:
\begin{itemize}[leftmargin=*]
    \item \textbf{RQ1:}  
    How does the proposed \textit{Cardiff} framework perform compared to state-of-the-art trajectory synthesis baselines?
    \item \textbf{RQ2:}  
    What are the contributions of the key technical components within \textit{Cardiff} to the overall performance?
    \item \textbf{RQ3:}  
    To what extent can the proposed model generate trajectories controllably under different conditional inputs? 
    \item \textbf{RQ4:}  
    How effective is \textit{Cardiff} at preserving privacy while maintaining utility?
\end{itemize}

\subsection{Evaluation Setups}
\label{sec:Evaluation Setups}
\subsubsection{Dataset Description}
To evaluate our designs, we conduct experiments based on three large-scale real-world trajectory datasets collected from Singapore, Chengdu, and Porto. 
The Singapore dataset is a part of the Veraset Dataset~\cite{veraset2024}, which comprises two months of human mobility data collected by mobile phones. 
The Chengdu dataset~\cite{didi2017gaia} consists of one month of trajectory data collected during November 2019. 
The Porto dataset~\cite{pkdd-15-taxi-trip-time-prediction-ii} contains trajectory data for 442 taxis from January 2013 to June 2014. 

In addition, we obtained detailed road network information for the above three cities from Openstreetmap~\cite{osm}. The road network features include road types, road lengths, directionality (i.e., whether the roads are unidirectional), as well as detailed information on location nodes and edges, specified by latitude and longitude coordinates. 
Table~\ref{tab:dataset} provides the statistical information on the datasets for the model training and validation.

\begin{table}[htbp]
\centering
\caption{Statistics of trajectory datasets.}
\label{tab:dataset}
\resizebox{0.95\linewidth}{!}{
\begin{tabular}{c|c|c|c}
\hline
Fields & Singapore & Chengdu & Porto \\ 
\hline
Latitude range & 1.22$\to$1.47 &30.65$\to$30.73  &  41.140$\to$ 41.186\\ 
Longitude range& 103.59$\to$104.05 &104.04 $\to$104.13 &  -8.690$\to$-8.550\\ 
\# of trajectory & 636K   & 463K & 1.4M \\ 
\# of road segments & 10871 & 6180  &15167 \\ 
\hline
\end{tabular}
}
\end{table}

\subsubsection{Implementation Details}
We implement our \textit{Cardiff} and other baselines with Pytorch 2.4.1, Python 3.9. 
All methods are trained with three Tesla L40s GPUs (each with 46GB of memory) to guarantee the fairness of performance comparison. 
We use the Adam Optimizer~\cite{kingma2014adam} with a learning rate of 1e-4.

The detailed settings of the denoising networks in our cascaded diffusion structure are summarized in Table~\ref{tab:parameters}.
For the overall hyperparameter configurations of \textit{Cardiff}, we set the number of diffusion steps to 1000 and adopt a linear noise scheduler for both the segment-level and fine-grained GPS-level diffusion.
In the first-stage discrete segment-level denoising network, the depth of the DiT layers ranges from 8 to 12, depending on the complexity of the trajectory data distribution. The input latent length is set to 32, with a feature dimension of 128.
For the second-stage continuous GPS-level denoising network, we adopt a U-Net architecture with the spatial-transformer attention mechanism. 
The GPS-level denoising network consists of four sampling layers, where the channel sizes are 128, 128, 256, and 256, respectively.
\begin{table}[ht]
\centering
\caption{Hyperparameters of the Cascaded Diffusion Model\label{tab:parameters}}
\fontsize{8}{12}\selectfont
\resizebox{0.95\linewidth}{!}{
\begin{tabular}{ll| ll}
\toprule
\multicolumn{4}{c}{\textbf{Hyperparameters of \textit{Cardiff}}} \\
\cline{1-4}
Diffusion steps     & 1000 & Latent dimension        & 128       \\
Noise schedule          & Linear    & Learning rate       & 1e-4 \\
Optimizer               & Adam      & Batch size          & 512 \\
\cline{1-4}
\midrule
\multicolumn{2}{c}{\textbf{Hyperparameters of segment-level $\epsilon_{\theta_r}$}} & \multicolumn{2}{c}{\textbf{Hyperparameters of GPS-level $\epsilon_{\theta_g}$}} \\
\cline{1-2} \cline{3-4}
DiT Layers depth     & 8 $\sim$ 12 & down/up Sampling Layers     & 4 \\
Hidden size             & 256             & channels     & (128,128,256,256) \\
Attention heads         & 4               & Attention heads & 4 \\
input length         & 32               & input length & 64 $\sim$ 256 \\
input dimension      & 128               & condition dimension & 128 \\
\cline{1-4}
\bottomrule
\end{tabular}
}
\end{table}

\subsubsection{Metrics}
To ensure the fidelity of the synthetic trajectory data, it is crucial to evaluate the similarity between the generated and real datasets.
In this paper, we adopt the Jensen-Shannon Divergence (JSD)~\cite{lin1991divergence} to quantify the difference between the real trajectory data distribution and the synthetic trajectory distribution, 
which is denoted as 
\begin{equation}
\operatorname{JSD}(p, q) = \frac{1}{2}\operatorname{KL}\left(p \,\bigg\|\, \frac{p+q}{2}\right)
+ \frac{1}{2}\operatorname{KL}\left(q \,\bigg\|\, \frac{p+q}{2}\right),
\end{equation}
where $\operatorname{KL}(\cdot)$ is the Kullback-Leibler divergence~\cite{kullback1951information} between two distributions. For $\operatorname{JSD}(p, q)$, a smaller value indicates that the two distributions are more similar.
For each dataset, 
we first utilize the \textit{Cardiff} and the baselines to sample 2000 synthetic trajectories.  
Then, based on Jensen-Shannon Divergence (JSD), we follow the metrics in previous trajectory synthesizing works~\cite{du2023ldptrace,gursoy2018utility,zhu2023difftraj} and utilize the following three metrics for performance comparisons between real and synthetic distributions. 

\begin{itemize}[leftmargin=*]
    \item JSD-SD: which means the Jensen-Shannon Divergence of the spatial distribution between synthetic trajectories and real trajectories. JSD-SD is computed by partitioning the city into 30 $\times$ 30 grids, counting the distribution of generated trajectory points within each grid cell, and then calculating the divergence between the two spatial distributions. 
    \item JSD-LD: which means the Jensen-Shannon Divergence of lengths between real trajectory and synthetic trajectory data. 
    \item JSD-trip: which means the Jensen-Shannon Divergence of the origin and destination grid distribution of the trajectory between real trajectory data and synthetic trajectory data.
\end{itemize}

\subsubsection{Baselines}

To evaluate the effectiveness of our proposed method, we compare it with the following four learning-based state-of-the-art trajectory synthesizing models. 

\begin{itemize}[leftmargin=*]
    \item TrajVAE~\cite{chen2021trajvae,kingma2013auto}: TrajVAE utilizes an LSTM combined with a variational autoencoder (VAE) to learn trajectory representations and then employs a decoder to reconstruct the corresponding GPS points.
    \item CNN-TrajGAN~\cite{merhi2024synthetic}: CNN-TrajGAN utilizes a convolutional neural network-based GAN to capture the spatial distribution and generate trajectories to balance the utility and privacy.  
    \item DiffTraj~\cite{zhu2023difftraj}: DiffTraj is a diffusion model-based trajectory synthesizing framework, which utilizes UNet as the denoising network to generate GPS-level trajectory directly.
    \item ControlTraj~\cite{zhu2024controltraj}: ControlTraj is an enhanced version of DiffTraj that incorporates road network trajectories as conditional supervisory signals to adhere to the road network topology, which requires providing road network-level trajectories as the conditional signal during the generation phase.
\end{itemize}

\subsubsection{Variations of Cardiff} We also design three variations of the proposed \textit{Cardiff} to conduct ablation studies and evaluate the effectiveness of different modules in our framework.
\begin{itemize}[leftmargin=*]
\item \textit{Cardiff} $w/o$ c: \textit{Cardiff} $w/o$ c modifies the trajectory autoencoder to a standard version without latent compression and external coordinates embeddings, which is used to assess the importance of compressed encoding design and to verify the impacts of the latent quality on final diffusion performances.
\item \textit{Cardiff} $w/o$ p: \textit{Cardiff} $w/o$ p removes the physics-based spatial validity loss term during first-stage segment-level latent diffusion training, which is utilized to evaluate the contribution of physical constraints.
\item \textit{Cardiff} $w/o$ aug: this variation means removing the low-level noise augmentation mechanism when performing fine-grained GPS-level denoising network training. \textit{Cardiff} $w/o$ aug directly utilizes first-level latent $z_0$ as conditions.
\end{itemize}

\subsection{Experimental Results}
\label{sec:results}

\begin{table*}[h]
    \centering
    \fontsize{11}{15}\selectfont
    \caption{Performance comparisons on three datasets. \textbf{Bold} and \underline{underline} indicate the best and the second-best result.}
    \label{tab:main-performace}
    \resizebox{0.88\textwidth}{!}{
    \begin{tabular}{lccc|ccc|ccc}
    \toprule
    \hline
        \multirow{2}*{Method} &
        \multicolumn{3}{c}{Singapore}& 
        \multicolumn{3}{c}{Chengdu}&
        \multicolumn{3}{c}{Porto}\\
        \cline{2-4}\cline{5-7}\cline{8-10}
        & JSD-SD $\downarrow$ &JSD-LD $\downarrow$ &JSD-trip $\downarrow$ 
        & JSD-SD $\downarrow$ &JSD-LD $\downarrow$ &JSD-trip $\downarrow$
        & JSD-SD $\downarrow$  &JSD-LD $\downarrow$ &JSD-trip $\downarrow$ \\
        \hline
        {VAE} &  0.0879& 0.0718 &0.0768 &0.0831  & 0.0881 &0.1895& 0.1334 & 0.0445 &0.3654     \\  
        {CNN-TrajGAN}& 0.0817 & 0.0707 & 0.0757& 0.0998 & 0.1106&0.1509 &0.1785  & 0.0350  &0.3065\\ 
        {DiffTraj} & 0.0285 & 0.0204 & 0.0343&  0.0239 & 0.0147 &0.0922& 0.0572 &  0.0205&0.1425 \\
        {ControlTraj} & 0.0287 & 0.0146 &0.0245& 0.0204 & 0.0170 &0.0762&0.0435  & \underline{0.0175} &0.1520 \\
        {Cardiff $w/o$ c} &\underline{0.0066}  &\underline{0.0106}  &\underline{0.0179}& 0.0129 &0.0104  &0.0358& 0.0358 & 0.0264&0.1655  \\ 
        {Cardiff $w/o$ p} & 0.0068 & 0.0130 &0.0204& 0.0156  & \textbf{0.0038} &0.0330& 0.0436 & 0.0214 &\underline{0.1280} \\
        {Cardiff $w/o$ aug} & 0.0081 & 0.0115 &0.0238& \underline{0.0128} & \underline{0.0063} &\underline{0.0326}& \underline{0.0321} & 0.0319 &0.1358 \\ 
        \textbf{Cardiff} & \textbf{0.0043} & \textbf{0.0089} &\textbf{0.0164}& \textbf{0.0120} & 0.0067 &\textbf{0.0320}& \textbf{0.0275}  &\textbf{0.0155} &\textbf{0.1170}\\ 
        \hline
        \bottomrule
    \end{tabular}
    }
\end{table*}

\subsubsection{Main Performance (\textbf{RQ1})}

Since the proposed cascaded hybrid trajectory synthesis aims to generate high-fidelity trajectories sharing the same distribution as the real-world dataset, 
we first evaluate the overall distribution similarity performance between \textit{Cardiff} and other baselines.
We utilize the trained \textit{Cardiff} as well as all baselines to generate the same number of synthetic trajectories as the synthetic test dataset. 
We then calculate the JSD-SD, JSD-LD, and JSD-trip metrics to compare the distributions between the synthetic datasets and the real trajectory dataset.
Table~\ref{tab:main-performace} shows the similarity performance comparisons.
Overall, the proposed \textit{Cardiff} achieves better performance than baselines. 
In the Singapore Veraset dataset, Cardiff achieves notably low distribution divergence, offering 0.0043 in JSD-SD, 0.0089 in JSD-LD metrics, and 0.0164 in JSD-trip, which indicates that the generated data is highly realistic. 
On the Porto taxi dataset and the Chengdu ridesharing dataset, Cardiff also attains the lowest JSD-SD and JSD-trip metrics, further demonstrating its ability to produce data that closely match real-world patterns. 

Particularly, we have the following analysis findings: (i) the diffusion-based trajectory generation models consistently outperform the GAN and VAE-based baselines. This is largely attributed to the gradual denoising process in diffusion models, which enables more accurate learning of complex data distributions and provides a more stable sampling procedure compared to GANs~\cite{goodfellow2014gan} and VAEs~\cite{kingma2013auto}. (ii) \textit{Cardiff} surpasses both DiffTraj~\cite{zhu2023difftraj}, which applies single-stage diffusion generation, and Controltraj~\cite{zhu2024controltraj}, which incorporates road network information into single-stage diffusion models. The superior performance of our cascaded framework stems from its ability to simultaneously capture both segment-level and fine-grained GPS-level trajectory distributions more effectively during cascaded training.

\subsubsection{Ablation Studies (\textbf{RQ2})}
As is shown in Table~\ref{tab:main-performace}, we conduct ablation analysis by comparing \textit{Cardiff} with three variations, i.e., \textit{Cardiff} without the latent compression layers in the trajectory autoencoder (\textit{Cardiff}  $w/o$ c), 
\textit{Cardiff} without spatial physical constraints (\textit{Cardiff} $w/o$ p), and \textit{Cardiff} without noise augmentation conditioning mechanisms (\textit{Cardiff} $w/o$ aug). 

From the results shown in the Table~\ref{tab:main-performace}, 
we observe that 
(i) for the JSD-SD metric, \textit{Cardiff} has better performance with the lowest spatial distribution divergence compared to the three variations in all three datasets, which verifies the effectiveness of our latent compression, spatial physical constraints, and noise augmentation conditioning design in maintaining the overall spatial distribution. 
(ii) For the JSD-LD metric, on the Singapore and Porto datasets, \textit{Cardiff} performs better than several variant models. On the Chengdu dataset, however, its performance is slightly worse than the two variants, possibly because the physical constraints introduced in the second-stage fine-grained generation impose excessive restrictions that reduce the model’s flexibility.

\begin{figure}[htbp]
    \centering
    \subfloat[Physical validity weighting]{%
        \includegraphics[width=0.49\linewidth]{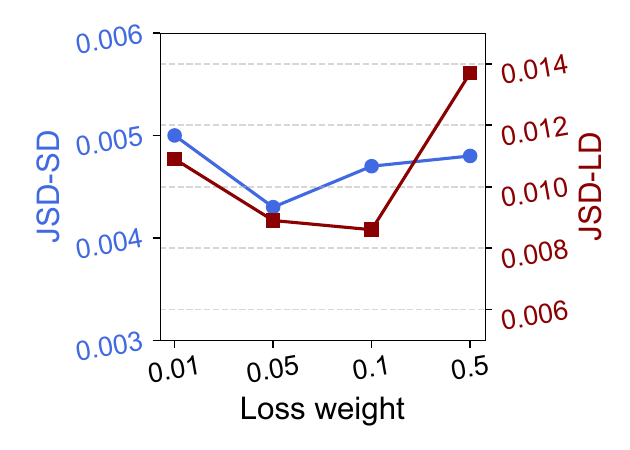}
    }
    \subfloat[Sampling interval steps]{%
        \includegraphics[width=0.49\linewidth]{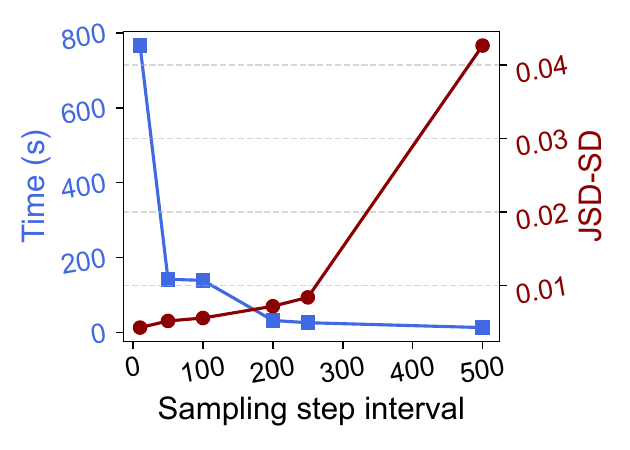}
    }
    \caption{Parameter sensitivity analysis on Singapore dataset: (a) Physical loss weighting during training, (b) Sampling interval setting for accelerated inference.}
    \label{fig:parameter-sensitivity}
\end{figure}

\subsubsection{Parameter Sensitivity Analysis}

We also conduct experiments to evaluate the impact of two key hyperparameters on generation performance, i.e., (i) the weight $\lambda_{p}$ of physical validity during training and (ii) the sampling interval during inference.

\noindent\textbf{Impact of physical validity weights:} 
As introduced in Sec.~\ref{sec:first-level-Cascaded-design}, the physical validity loss enforces segment continuity in the generated trajectories, preventing unrealistic jumps across the road network. As shown in Fig.~\ref{fig:parameter-sensitivity}(a), we evaluate the impact of the validity loss on the Singapore dataset by setting $\lambda_{p}$ (in Equation~\ref{equ:first-level-loss}) to 0.01, 0.05, 0.1, and 0.5. 
We observe that JSD-SD remains relatively stable across different weights, likely due to the robustness of the \textit{Cardiff} in handling the noise from first-stage segment-level conditioning. The JSD-LD metric increases significantly with higher weights, indicating that excessive emphasis on physical constraints may interfere with the diffusion model’s reconstruction objectives. Based on this observation, we select 0.05 as the default weight for the physical loss on the Singapore dataset. For the Chengdu and Porto datasets, the weights are set to 0.05 and 0.01, respectively, according to the same set of hyperparameter experiments.

\noindent\textbf{Impact of sampling interval steps:}
Diffusion models suffer from slow sampling speeds due to a long reverse Markov denoising chain~\cite{ho2020denoising} (in our scenarios, we set it to 1000 steps). 
To address this, DDIM~\cite{song2020denoising} was proposed as a non-Markovian deterministic variant of DDPM that enables accelerated sampling by skipping steps in the diffusion trajectory while preserving sample quality. 

Motivated by this, we evaluate the effect of the sampling interval parameter on both generation quality and efficiency. As shown in Fig.~\ref{fig:parameter-sensitivity}(b), we test various interval settings on the Singapore dataset. When using an interval of 50 (corresponding to $\frac{1000}{50}= 20$ denoising steps), the model generates 1,024 trajectories with minimal performance degradation, while reducing the sampling time by nearly six times compared to using an interval of 10. Based on the trade-off between performance and efficiency, we adopt an interval of 50 as the default setting for the sampling phase in our \textit{Cardiff} model.

\begin{figure}[h]
  \centering
  \begin{subfigure}[b]{0.32\linewidth}
    \includegraphics[width=\linewidth]{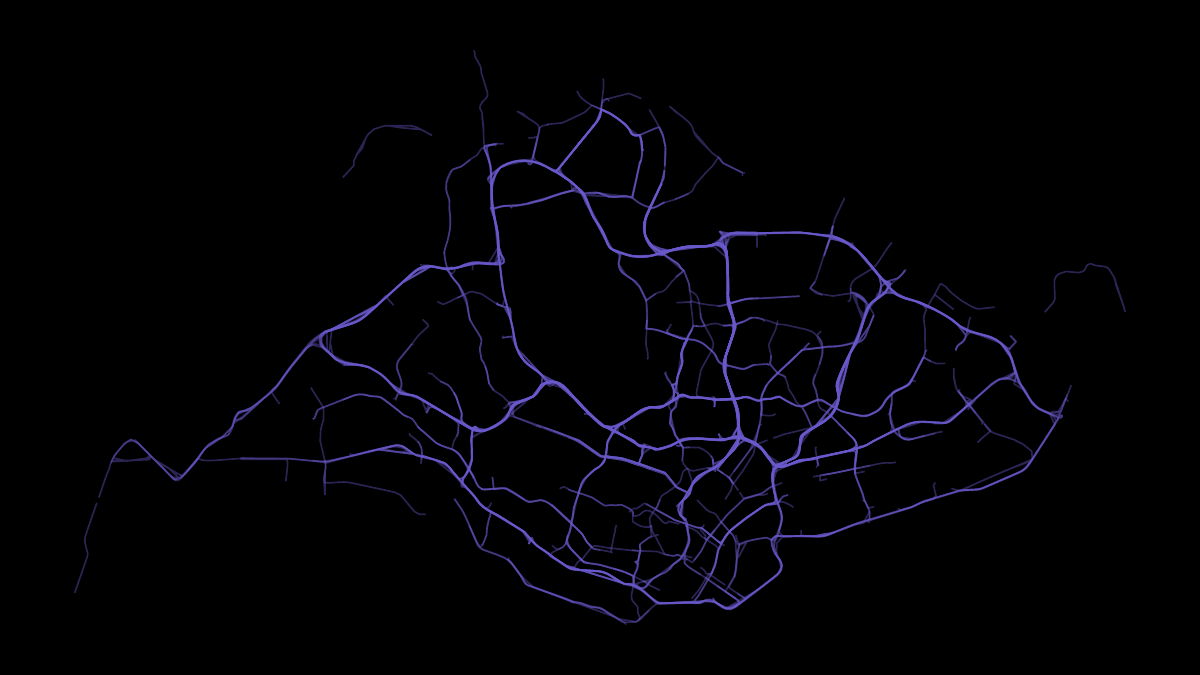}
    \caption{Real}
    \label{fig:sgreal}
  \end{subfigure}
  \hfill
  \begin{subfigure}[b]{0.32\linewidth}
    \includegraphics[width=\linewidth]{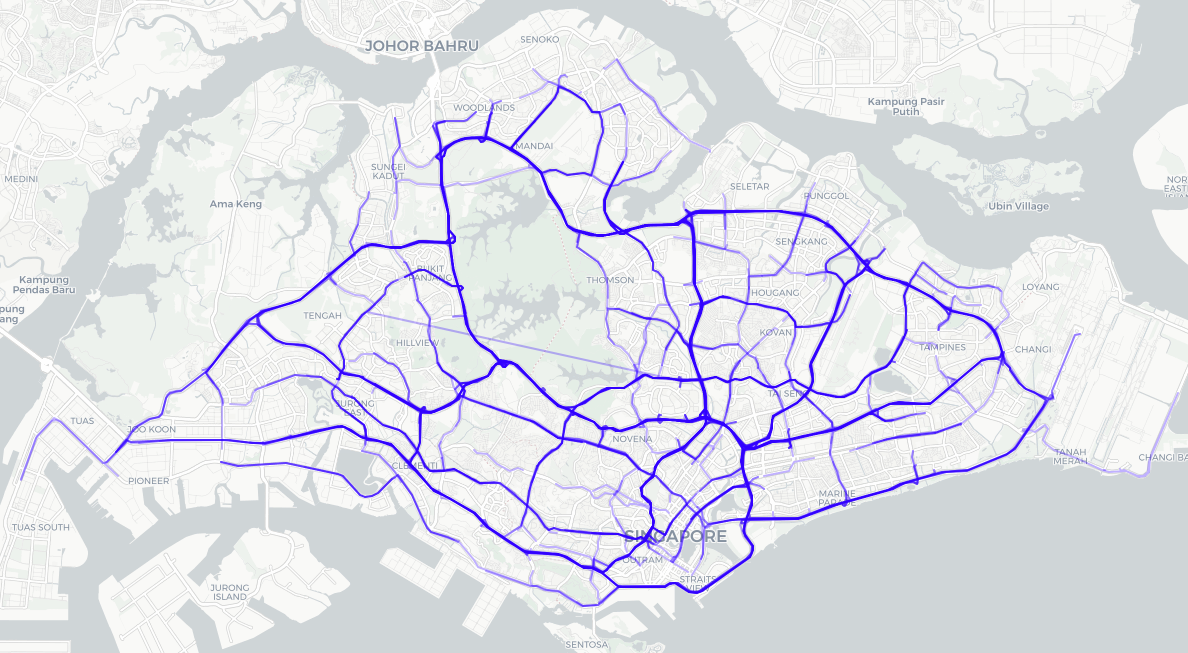}
    \caption{Segment-level}
    \label{fig:sgsynthetic}
  \end{subfigure}
  \hfill
  \begin{subfigure}[b]{0.32\linewidth}
    \includegraphics[width=\linewidth]{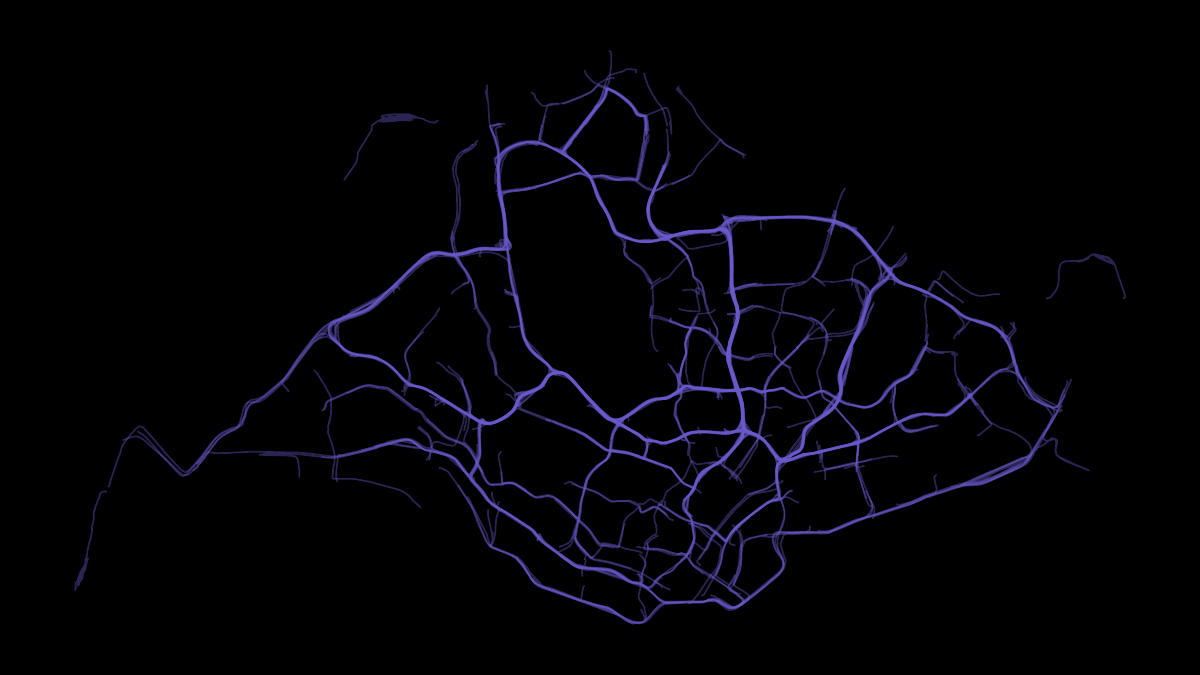}
    \caption{GPS-level}
    \label{fig:sgsegment}
  \end{subfigure}
  \caption{Results visualization of Singapore dataset: (a) real trajectories, (b) segment-level synthetic trajectories, and (c) fine-grained synthetic trajectories.}
  \label{fig:sg_trajectory_comparison}
\end{figure}

\begin{figure}[h]
  \centering
  \begin{subfigure}[b]{0.29\linewidth}
    \includegraphics[width=\linewidth]{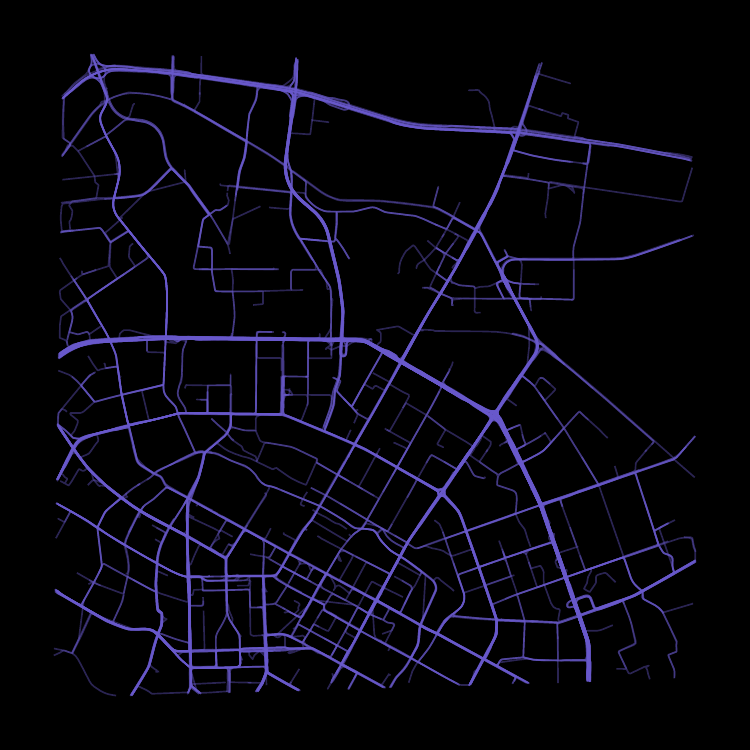}
    \caption{Real}
    \label{fig:cdreal}
  \end{subfigure}
  \hfill
  \begin{subfigure}[b]{0.29\linewidth}
    \includegraphics[width=\linewidth]{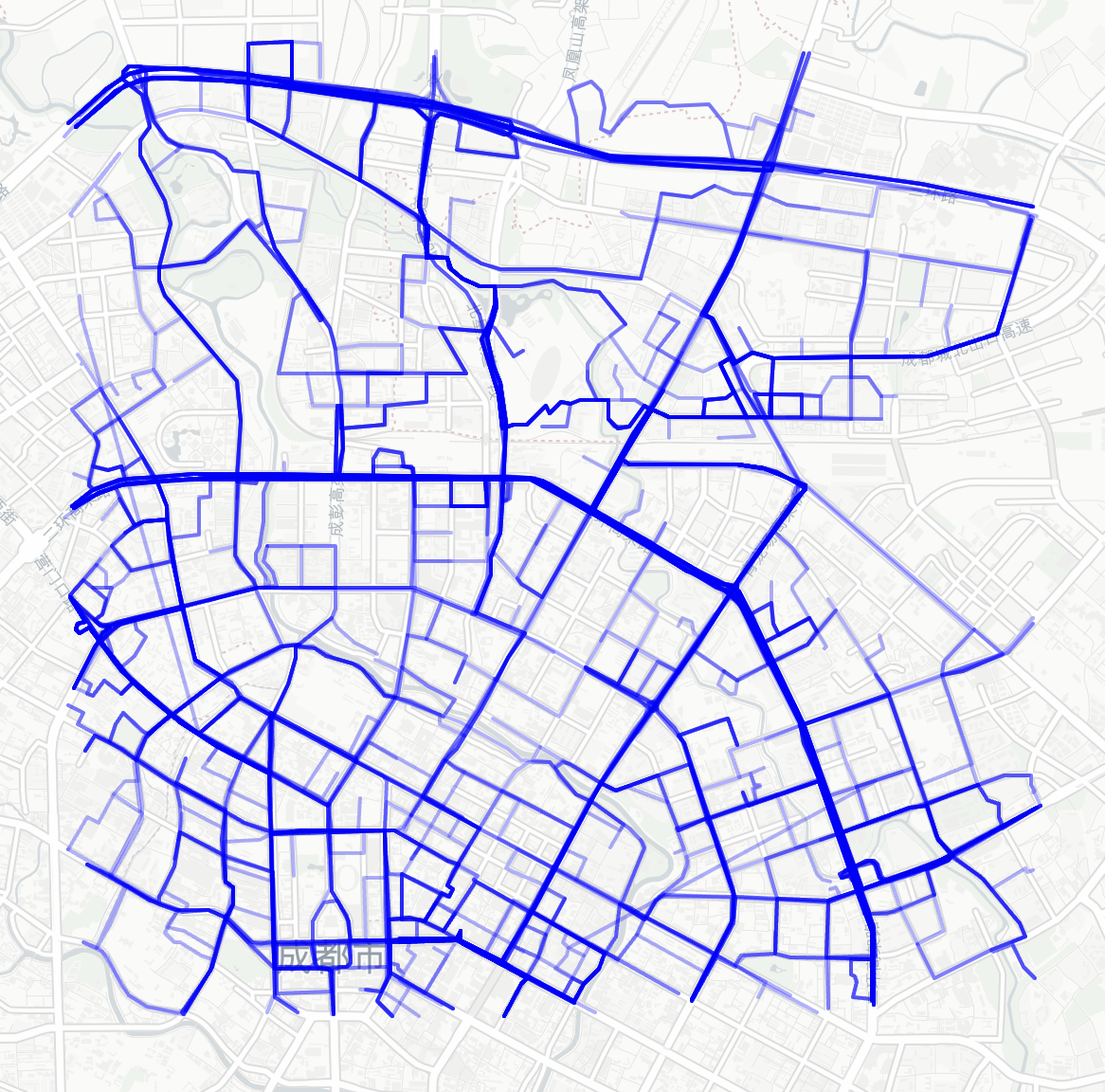}
    \caption{Segment-level}
    \label{fig:cdsynthetic}
  \end{subfigure}
  \hfill
  \begin{subfigure}[b]{0.29\linewidth}
    \includegraphics[width=\linewidth]{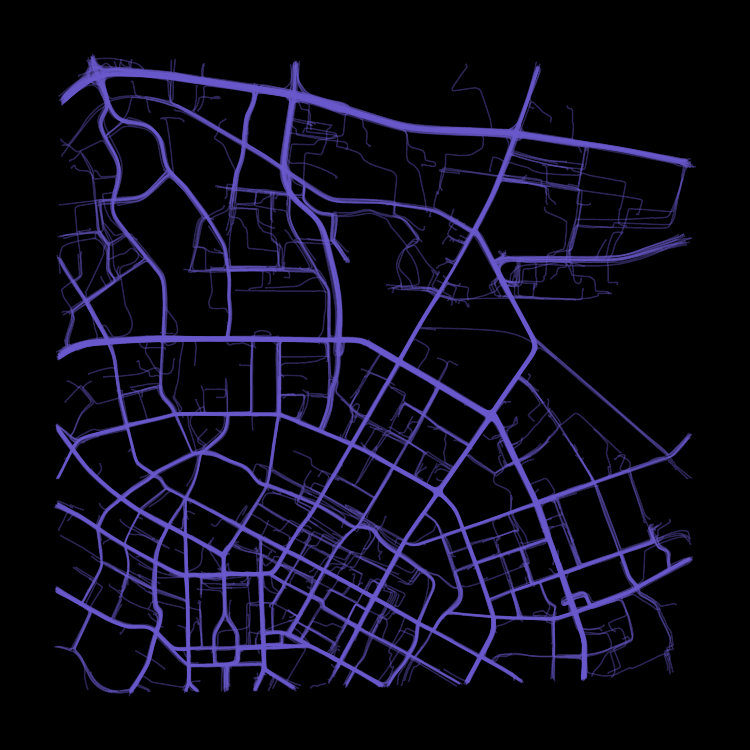}
    \caption{GPS-level}
    \label{fig:cdsegment}
  \end{subfigure}
  \caption{Results visualization of Chengdu dataset: (a) real trajectories, (b) segment-level synthetic trajectories, and (c) fine-grained synthetic trajectories.}
  \label{fig:cd_trajectory_comparison}
\end{figure}

\begin{figure}[h]
  \centering
  \begin{subfigure}[b]{0.31\linewidth}
    \includegraphics[width=\linewidth]{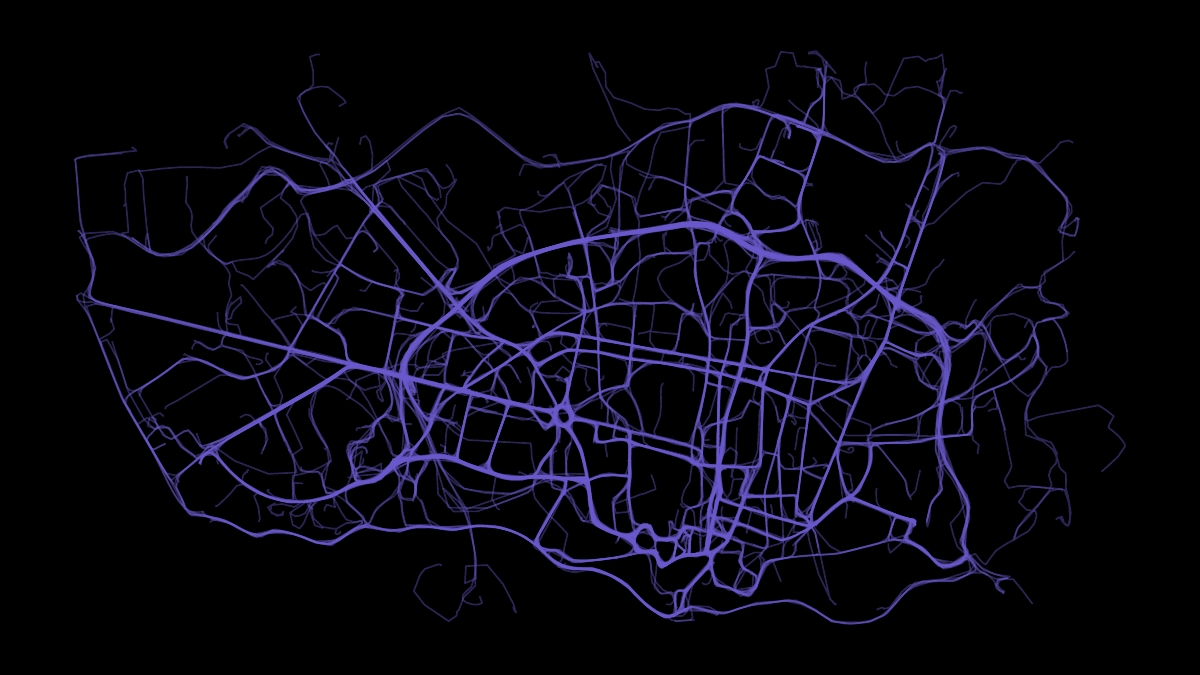}
    \caption{Real}
    \label{fig:poreal}
  \end{subfigure}
  \hfill
  \begin{subfigure}[b]{0.325\linewidth}
    \includegraphics[width=\linewidth]{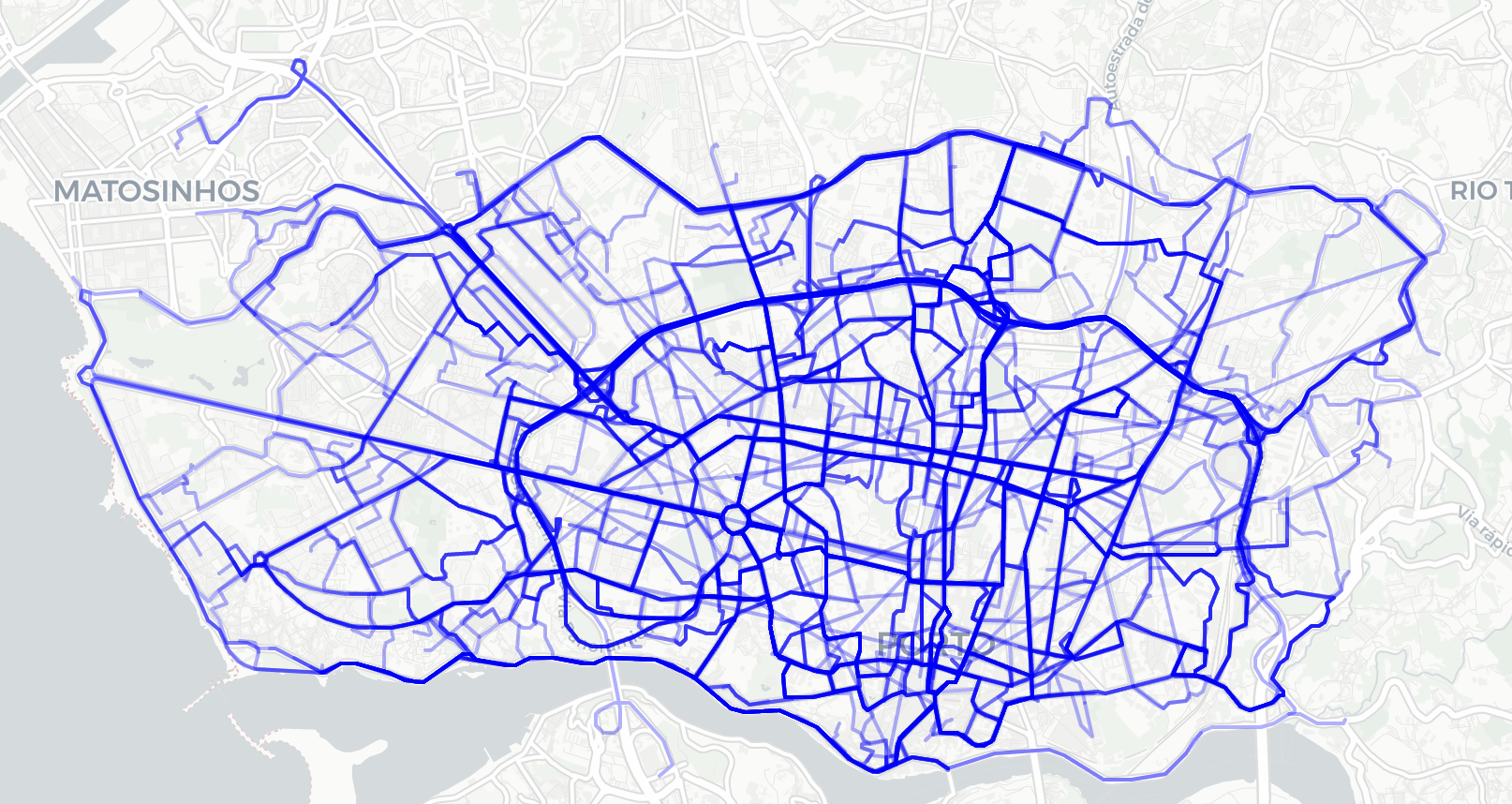}
    \caption{Segment-level}
    \label{fig:posynthetic}
  \end{subfigure}
  \hfill
  \begin{subfigure}[b]{0.31\linewidth}
    \includegraphics[width=\linewidth]{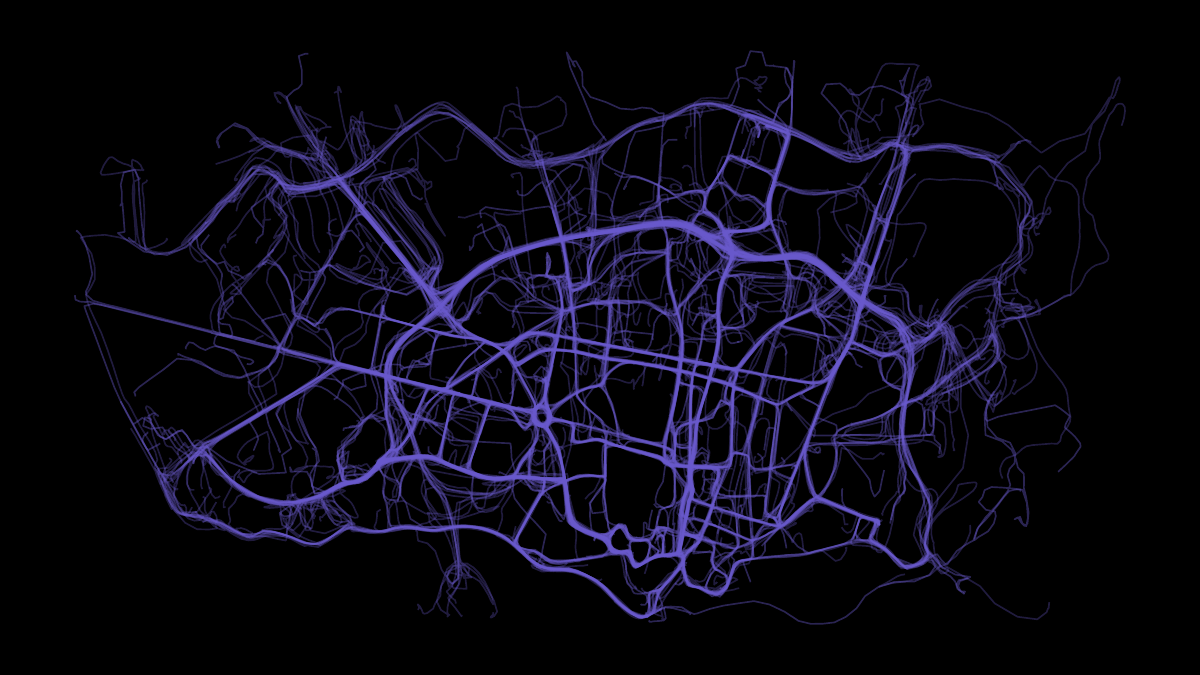}
    \caption{GPS-level}
    \label{fig:posegment}
  \end{subfigure}
  \caption{Results visualization of Porto dataset: (a) real trajectories, (b) segment-level synthetic trajectories, and (c) fine-grained synthetic trajectories.}
  \label{fig:po_trajectory_comparison}
\end{figure}

\subsubsection{Trajectory Visualization Comparisons}
To more intuitively compare and evaluate the similarity between the generated results and the real data, we generate and select an equal number (1024) of synthetic and real trajectories for visualization comparisons.

\noindent\textbf{Spatial visualization comparisons:}
To facilitate a more intuitive comparison between the synthesized trajectories and real-world data, and to validate the high fidelity of our generation results, we additionally conduct spatial distribution visualizations based on aggregated trajectories. 
Fig.~\ref{fig:sg_trajectory_comparison}, Fig.~\ref{fig:cd_trajectory_comparison}, and Fig.~\ref{fig:po_trajectory_comparison} present the spatial visualizations of trajectory data across three cities, including the spatial distribution of real trajectories, the generated distribution of segment-level trajectories over the road network, and the spatial distribution of fine-grained GPS-level trajectories.
By comparison, we observe that the segment-level trajectories consistently align with the underlying road network. Moreover, the fine-grained generated trajectories closely adhere to the road structure and exhibit a high degree of realism compared to the real trajectories.
Combined with the quantitative results reported in Table 3, these observations demonstrate that our \textit{Cardiff} model effectively preserves the spatial distribution characteristics of the generated trajectories.

\begin{figure}[htbp]
    \centering
    \subfloat[Stepwise Distance]{%
        \includegraphics[width=0.48\linewidth]{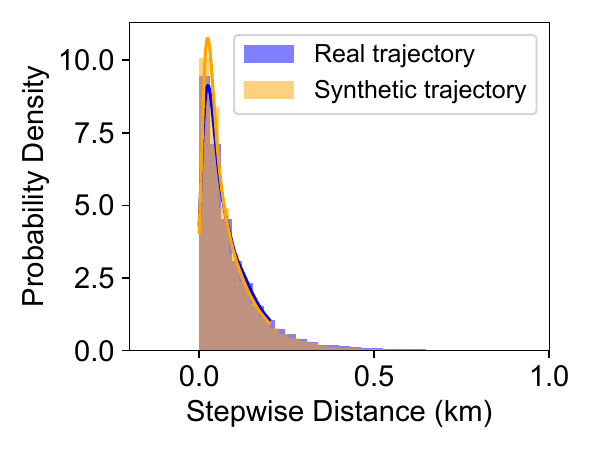}
    }
    \subfloat[Trip Length]{%
        \includegraphics[width=0.48\linewidth]{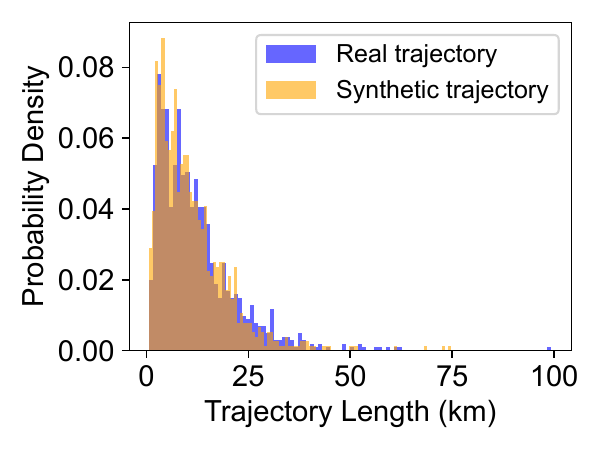}
    }
    \caption{Trajectory length distributions in Singapore.}
    \label{fig:singapore_distributions}
\end{figure}

\begin{figure}[htbp]
    \centering
    \subfloat[Stepwise Distance]{%
        \includegraphics[width=0.48\linewidth]{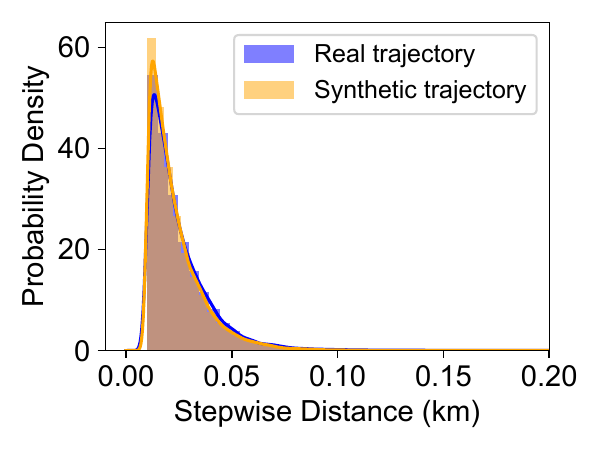}
    }
    \subfloat[Trip Length]{%
        \includegraphics[width=0.48\linewidth]{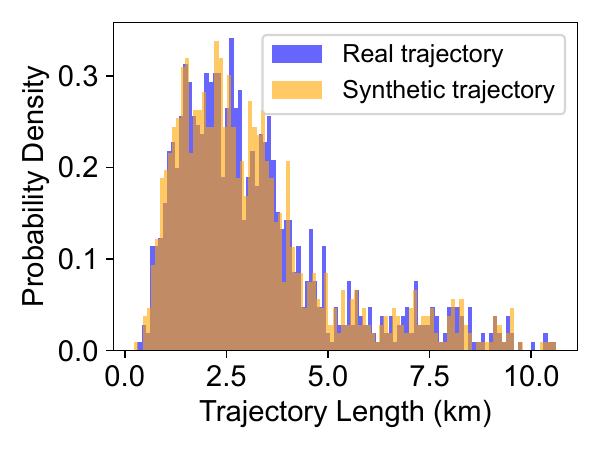}
    }
    \caption{Trajectory length distributions in Chengdu.}
    \label{fig:chengdu_distributions}
\end{figure}

\begin{figure}[htbp]
    \centering
    \subfloat[Stepwise Distance]{%
        \includegraphics[width=0.48\linewidth]{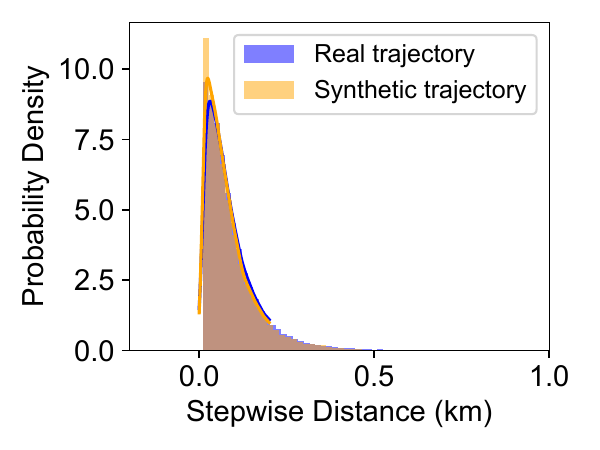}
    }
    \subfloat[Trip Length]{%
        \includegraphics[width=0.48\linewidth]{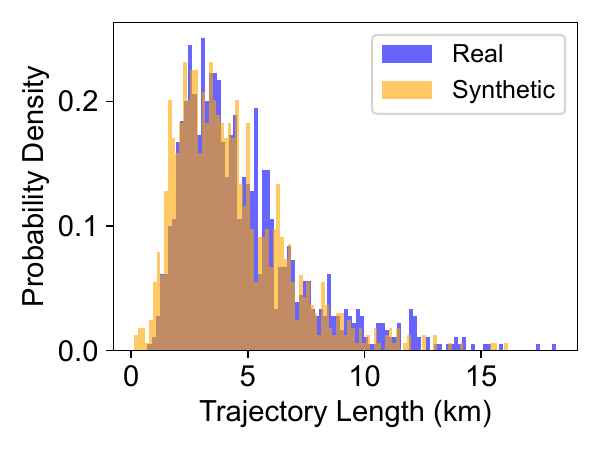}
    }
    \caption{Trajectory length distributions in Porto.}
    \label{fig:porto_distributions}
\end{figure}

\noindent\textbf{Individual-level trajectory length distributions:}
In terms of individual-level trajectory length distributions, we compare the differences between real and synthetic data to both step-wise point-to-point distances and overall trajectory lengths.
As shown in Fig.~\ref{fig:singapore_distributions}, Fig.~\ref{fig:chengdu_distributions}, and Fig.~\ref{fig:porto_distributions}, the results of \textit{Cardiff} closely match the real data distributions.
Furthermore, while the trajectory length statistics exhibit different patterns across the three city datasets, our method consistently achieves high-quality fitting to the underlying distributions.

\subsubsection{Conditional Generation Evaluation (\textbf{RQ3})}
\begin{figure}[htbp]
  \centering
  \begin{subfigure}[b]{0.3\linewidth}
    \includegraphics[width=\linewidth]{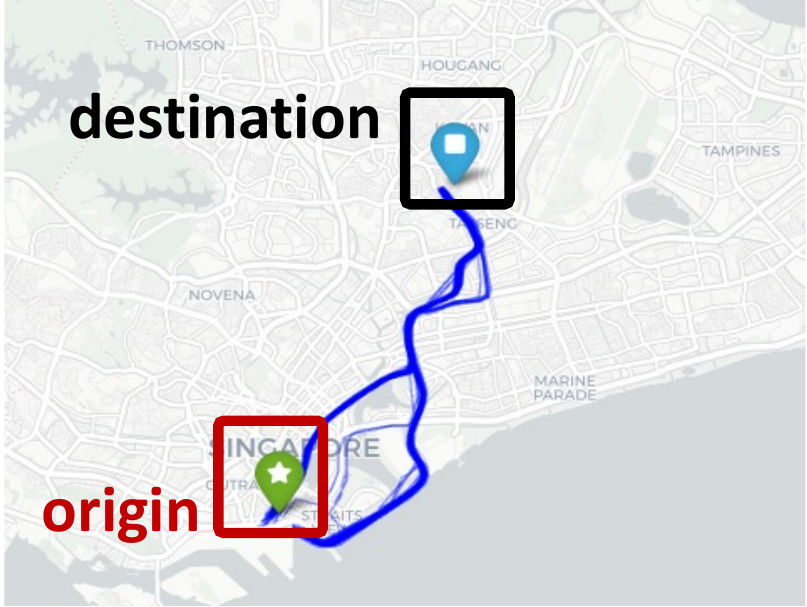}
    \caption{}
    \label{fig:trajectory_conditional-a}
  \end{subfigure}
  \hfill
  \begin{subfigure}[b]{0.3\linewidth}
    \includegraphics[width=\linewidth]{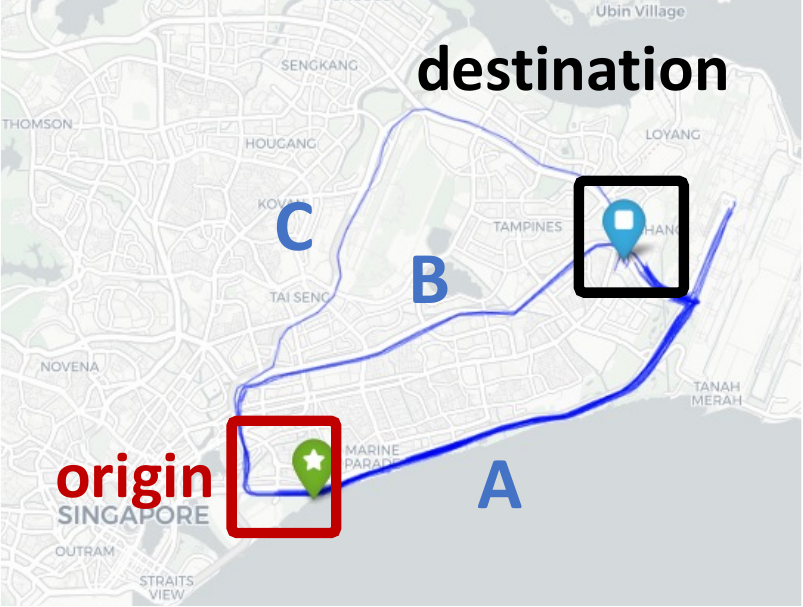}
    \caption{}
    \label{fig:trajectory_conditional-b}
  \end{subfigure}
  \hfill
  \begin{subfigure}[b]{0.3\linewidth}
    \includegraphics[width=\linewidth]{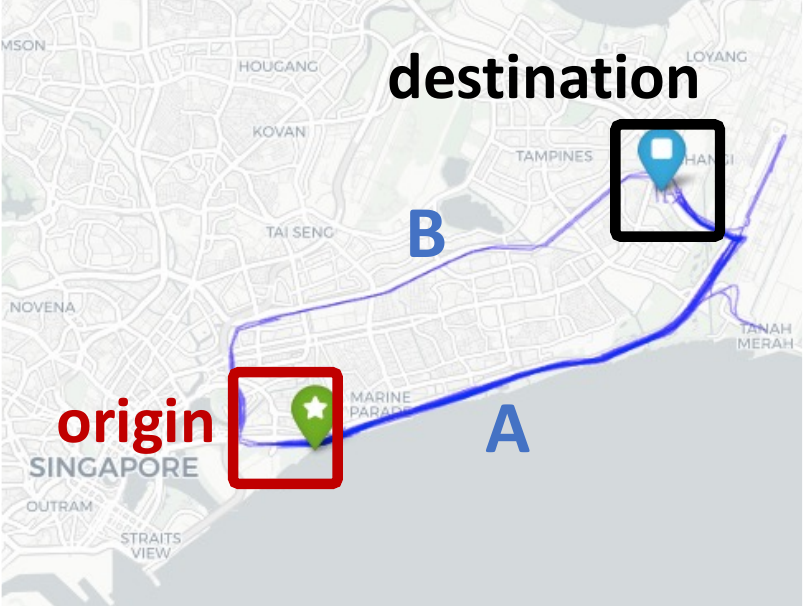}
    \caption{}
    \label{fig:trajectory_conditional-c}
  \end{subfigure}
  \caption{Visualization of conditional generation.}
  \label{fig:trajectory_conditional}
\end{figure}

We also evaluate the conditional generation capabilities of \textit{Cardiff}.
Fig.~\ref{fig:trajectory_conditional-a} and Fig.~\ref{fig:trajectory_conditional-b} illustrate the generation results given different origin and destination grids (with each city divided into small spatial grids of size 500 $\times$ 500 meters). It can be observed that the model accurately generates trajectories that strictly follow the specified start and end conditions.
Furthermore, comparing Fig.~\ref{fig:trajectory_conditional-b} and Fig.~\ref{fig:trajectory_conditional-c} where both share the same origin and destination grids but differ by an additional constraint on trajectory length, we observe that Fig.~\ref{fig:trajectory_conditional-c} successfully avoids generating the longer trajectory present in Fig.~\ref{fig:trajectory_conditional-b}, demonstrating the model’s ability to satisfy fine-grained conditional constraints.

\subsubsection{Utilities Evaluation of Downstream Tasks}
The above experiments have demonstrated that the proposed framework can generate high-fidelity fine-grained trajectory data, which preserves distributional characteristics comparable to real-world trajectories.
To further evaluate the practical utility of our synthetic data, 
following existing utility evaluation studies of synthetic trajectories~\cite{kapp2023generative,wei2024diff},
we choose next-location prediction~\cite{kapp2023generative} as the downstream application. 

\begin{table}[h]
    \centering
    \fontsize{10}{18}\selectfont
    \caption{Next segment prediction performance comparisons between different mixture rates of real data.}
    \label{tab:utility}
    \resizebox{0.9\linewidth}{!}{
    \begin{tabular}{lcccccc}
    \toprule
    \hline
        \multirow{2}*{Dataset} &
        \multicolumn{6}{c}{Accuracy with different ratio of real data }\\
        \cline{2-7}
        & 0\% &20\%   
        & 40\% &60\%  
        & 80\% &100\% \\
        \hline
        {Singapore} &  0.8882& 0.9021 &0.9066  & 0.9047 & 0.9065 & 0.9101      \\  
        {Chengdu}& 0.8224 & 0.8315 & 0.8373 & 0.8415 & 0.8453 & 0.8548  \\ 
        {Porto} &  0.8467& 0.8600 & 0.8637 & 0.8688 & 0.8727 & 0.8785  \\ 
        \hline
        \bottomrule
    \end{tabular}
    }
\end{table}

In the utility evaluation, we use the synthetic trajectories as the training set and real trajectories as the test set, aiming to assess the effectiveness of the generated data in supporting downstream tasks.
We also investigate the impact of incorporating different proportions of real data into the training set to examine how hybrid datasets influence the downstream tasks. 
As illustrated in Table~\ref{tab:utility}, 
the synthetic trajectories generated by \textit{Cardiff} contribute high accuracy on the next-position prediction task, demonstrating its effectiveness for downstream applications. 
On average, synthetic trajectories generated by Cardiff achieved similar performance compared to training on the real trajectory dataset.
Moreover, as the proportion of real data gradually increases from 0\% to 100\%, the prediction performance improves slightly accordingly.

\subsubsection{Privacy Evaluation and Privacy Guarantee (\textbf{RQ4})}
\label{sub-sec:privacy-evaluation}

Due to our cascaded diffusion design, our trajectory generation model offers enhanced flexibility in privacy preservation and evaluation. Specifically, we can perform differentially private gradient descent training at different stages. 
For instance, if the goal is to protect sensitive location information (such as home and workplace)~\cite{kapp2023generative}, we can apply differential privacy training (e.g., DP-SGD~\cite{abadi2016deep}) to the fine-grained diffusion model while keeping the coarse-grained stage under normal training. This strategy ensures the overall spatial distribution of trajectories while preserving the privacy of sensitive locations. Similarly, in scenarios involving coarse-grained privacy attacks, such as those targeting privacy-preserving for POI sequence recommendations~\cite{cai2024have}, we can inject differential noise solely into the training of the denoising model in the first stage. 

\begin{figure}[h] 
\centering 
\includegraphics[width=0.45\textwidth]{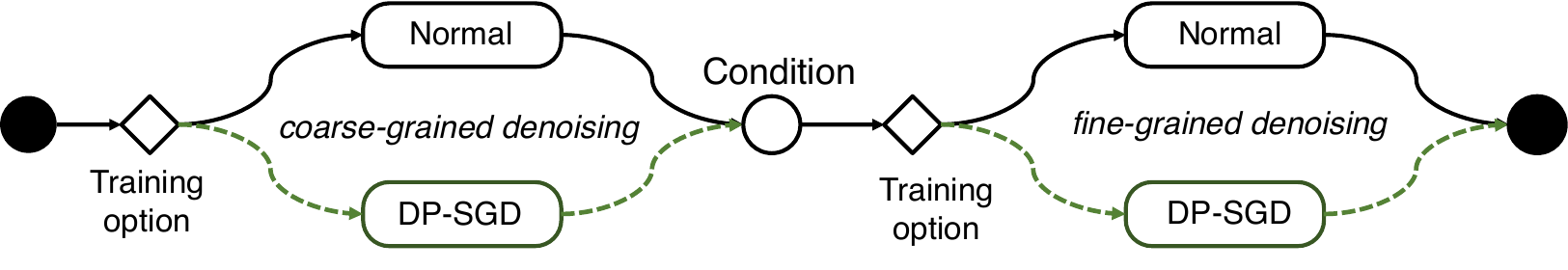} 
	\caption{Privacy-aware cascaded model training.\label{fig:cascaded-privacy}}
\end{figure}

To systematically evaluate the privacy and utility trade-offs, we design experiments under four different training settings (shown in Fig.~\ref{fig:cascaded-privacy}): (i) no differential privacy (DP) training (\textit{Cardiff} $w/o$ DP), (ii) applying DP training only during the first stage (\textit{Cardiff}-DP1), (iv) applying DP training only during the second stage (\textit{Cardiff}-DP2), and (v) applying DP training during both stages (\textit{Cardiff} $w/$ DP).
For the privacy evaluation, we employ the uniqueness test~\cite{de2013unique,yuan2024generating}, which measures the minimum distance between a generated trajectory and the closest real trajectory in the training database. 
In our setting, we choose the Hausdorff distance~\cite{huttenlocher1993comparing} as the uniqueness testing score for privacy measurement.
A low uniqueness score indicates a higher risk of privacy leakage, whereas a larger minimum distance suggests stronger privacy protection and less overfitting to individual real trajectories.

\begin{figure}[h]
    \centering
    \begin{minipage}[b]{0.45\linewidth}
        \centering
        \includegraphics[width=0.98\linewidth]{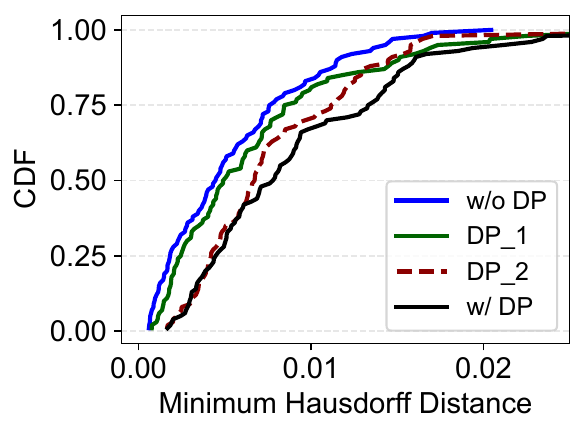}
        \captionof{figure}{CDF of the minimum Hausdorff distance to real data.\label{fig:privacy-evaluation}}
        \vspace{\baselineskip}
    \end{minipage}
\hfill
\begin{minipage}[b]{0.45\linewidth}
    \centering
    \fontsize{8}{18}\selectfont
    \resizebox{\linewidth}{!}{
    \begin{tabular}{lccc}
        \toprule
        \hline
        Model & JSD-SD & JSD-LD& JSD-Trip\\
        \hline
        $w/o$ DP & 0.0043 & 0.0089&0.0164 \\
        DP1 & 0.0052 & 0.0126& 0.0184\\
        DP2& 0.0246 & 0.0447 &0.0348\\
        $w/$ DP & 0.0223 & 0.0399&0.0416 \\
        \hline
        \bottomrule
    \end{tabular}
    }
    \vspace{\abovecaptionskip}
    \captionof{table}{Utility comparisons between $w/$ DP and $w/o$ DP training.\label{tab:privacy}}

\end{minipage}
\end{figure}

As shown in Fig.~\ref{fig:privacy-evaluation} and Table.~\ref{tab:privacy},
experimental results show that even without any DP training, the trajectories generated by \textit{Cardiff} do not achieve a zero minimum distance to any real trajectories, indicating that the generated data does not exactly replicate real-world trajectories.
Moreover, incorporating DP training in the first stage slightly improves privacy protection, while applying DP in the second stage introduces more significant perturbations to the generated data, offering stronger privacy guarantees.
In terms of utility, applying DP-SGD inevitably leads to a degradation in performance. However, we observe that when DP-SGD is applied only in the first stage, the reduction in utility is relatively minor, demonstrating that our model can effectively balance privacy protection with the preservation of data utility.

%% file: Sec5-relatedworks.tex
\section{Related Works}

\subsection{Trajectory Synthesizing}

Trajectory Synthesizing is an essential research problem in spatial-temporal data mining and smart cities~\cite{kapp2023generative,deng2024revisiting,takagi2024hrnet}, which not only benefits massive downstream tasks~\cite{zhang2025large} (e.g., urban planning and transportation management) but alleviates the privacy leakage concerns~\cite{du2023ldptrace,wang2023pategail} of using large-scale human mobility datasets~\cite{du2023ldptrace,zhang2023trajectory}. 
Existing trajectory synthesizing methods can be divided into 
(i) rule-based data mixing and augmenting methods~\cite{nergiz2008towards} 
and (ii) generative-based methods~\cite{kapp2023generative,long2023practical,zhu2023difftraj,chen2021trajvae}. 
Traditional rule-based synthesizing methods usually add random perturbations~\cite{nergiz2008towards}  to existing data sources, or mix different sources of data to obtain the new synthetic dataset, which will disrupt the original data distribution, thereby affecting the utility of synthetic trajectories.

With the rapid development of generative artificial intelligence, 
learning-based generative methods~\cite{cao2021generating,chen2021trajvae,shao2024beyond,wen2023diffstg}
have been widely adopted for trajectory synthesizing, 
which can be further divided into autoregressive-based and non-autoregressive-based studies. 
Autoregressive-based methods~\cite{chen2021trajvae,zhang2022factorized,shao2024beyond,rao2025seed} are typically built on RNN~\cite{schmidt2019recurrent}, LSTM~\cite{greff2016lstm}, or Transformer~\cite{vaswani2017attention} architectures. They discretize trajectories into token sequences and generate them in a left-to-right, step-by-step manner. These models show better training efficiency and fit well with discrete data, but suffer from constrained sampling and cumulative errors, making them less suitable for modeling continuous trajectory structures.
Non-autoregressive-based methods, including GAN-based~\cite{cao2021generating,jiang2023continuous,zhang2023dp,rao2020lstm} and diffusion-based approaches~\cite{wen2023diffstg,zhu2023difftraj,wei2024diff}, generate the entire trajectory holistically,
which better suits the generation of fine-grained, continuous data with complicated distributions, but suffers from lower training efficiency.
In this work, given the objective of synthesizing continuous and fine-grained trajectory data, we adopt diffusion models due to their inherent suitability for continuous data generation and propose a cascaded hybrid diffusion architecture to enhance training efficiency while preserving generation quality.

\subsection{Diffusion Models and Applications}
Diffusion models~\cite{ho2020denoising,nichol2021improved,saharia2022photorealistic,dhariwal2021diffusion_beat_gan, Karras2022edm} and score-based generative models~\cite{song2019generative} have achieved particular success in image~\cite{rombach2022high}, video~\cite{ho2022video}, time series~\cite{chen2023imdiffusion}, and other generation tasks, outperforming the previous generative adversarial networks and variational autoencoders~\cite{dhariwal2021diffusion_beat_gan}. 
Some early and widely used diffusion models, e.g., DDPM~\cite{ho2020denoising}, DDIM~\cite{song2020denoising}, and EDM~\cite{Karras2022edm} typically operate directly in pixel space to learn the distribution of images and then perform generation. 
Furthermore, related work on improving DDPM has largely enhanced the quality of generation from various perspectives. 
Latent diffusion models~\cite{rombach2022high} perform the diffusion process within a learned latent space by encoding high-dimensional data into a lower-dimensional latent representation, which accelerates both training and sampling while maintaining high-quality outputs.
Diffusion Transformer~\cite{peebles2023scalable} improves the image generation quality by replacing the Unet~\cite{ronneberger2015u} structure in traditional DDPM with transformer structures~\cite{vaswani2017attention}. 
Conditional diffusion models~\cite{nichol2021improved,ho2022classifier,saharia2022photorealistic} focus on imposing additional conditions and guidance to better guide the generation process, including class-based guidance~\cite{nichol2021improved}, classifier-free guidance~\cite{ho2022classifier}, and text2image generation~\cite{saharia2022photorealistic}.
Some works focus on conditional generation by performing image class-level guidance or class-free guidance.
In addition, there are also works~\cite{ho2022cascaded,dhariwal2021diffusion_beat_gan,saharia2022photorealistic} that explore improving diffusion models by using cascaded DDPM pipelines where low-resolution base diffusion models are
trained in parallel and generate the images from low-resolution to high-resolution progressively. 
The aforementioned models are all designed for continuous element generation, e.g., images or videos. 
Diffusion models have also been applied to discrete generation tasks~\cite{gong2022diffuseq} (e.g., language~\cite{lovelace2023latent}, path~\cite{shi2024graph}, and human motion~\cite{tevet2022human}) and improved the performance of these applications. 
 

%% file: Sec6-discussion.tex
\section{Discussion}
\noindent\textbf{Insights and Lessons Learned:}
Based on our studies and experimental results analysis, we learned the following lessons:
\begin{itemize}[leftmargin=*]
    \item \textit{Cascaded diffusion is well-suited for hierarchically structured mobility data.}  
    The hierarchy in mobility data naturally aligns with the coarse-to-fine design of cascaded diffusion. By progressively modeling the data distribution across two stages (i.e., segment-level and GPS-level), our model first generates plausible, road-network-consistent segment-level trajectories in the coarse stage, followed by refined, high-resolution GPS trajectories in the fine stage. This coarse-to-fine structure enables more efficient and robust trajectory synthesis.
    \item \textit{Learning a better latent representation is critical for latent trajectory diffusion learning.}  We observe that it is not necessary for the encoder and decoder to share an exactly symmetric architecture. 
    By incorporating additional features during the encoding stage, this model can retain sufficient information to accurately reconstruct the original trajectory and reduce the dimensionality of the latent space to enhance the training efficiency.
\end{itemize}

\noindent\textbf{Limitations and future works:}
Privacy guarantees remain a critical concern in trajectory generation. 
For our \textit{Cardiff} framework, the privacy evaluation results demonstrate that the cascaded design can support flexible and effective privacy guarantees for trajectory synthesis.  
However, our current approach does not yet fully exploit the inherent noise injection of diffusion models for coordination with differential privacy mechanisms. In the future, we will further explore how the diffusion noise inherent in denoising processes can be systematically combined with the noise applied in differential privacy mechanisms. We aim to explore the possibility of repurposing the diffusion noise to strengthen privacy guarantees in a more effective and efficient manner.

%% file: Sec7-conclusion.tex
\section{Conclusion}
In this paper, we introduce a coarse-to-fine cascaded hybrid diffusion model-based trajectory synthesizing framework for fine-grained and privacy-preserving mobility generation. 
Leveraging the hierarchical structure of urban mobility, we first decompose the trajectories into discrete road segment-level and continuous fine-grained GPS-level trajectories.  
Then, we integrate the discrete and continuous generation with a unified cascaded diffusion framework and generate the latent of discrete road segment sequences and fine-grained trajectories progressively.
Experimental results demonstrate the effectiveness of the proposed method for generating high-fidelity synthetic trajectories with flexible privacy guarantees.